\documentclass[a4paper,12pt]{article}
\usepackage{amsmath}
\usepackage{amsfonts}
\usepackage{authblk}
\usepackage{graphicx,epsfig}
\usepackage{float}
\usepackage{subcaption}  
\usepackage{amssymb}
\usepackage{color}

\numberwithin{equation}{section}
\title{Rindler horizons in the Schwarzschild spacetime}

\author{Kajol Paithankar\footnote{kajol.paithankar@cbs.ac.in} }
\author{Sanved Kolekar\footnote{sanved.kolekar@cbs.ac.in} \,}
\affil{ UM-DAE Centre for Excellence in Basic Sciences,\\  
Mumbai 400098, India}

\date{June 2019}
\begin{document}

\maketitle

\begin{abstract}

We investigate the past and future Rindler horizons for radial Rindler trajectories in the Schwarzschild spacetime. We assume the Rindler trajectory to be linearly uniformly accelerated (LUA) throughout its motion, in the sense of the curved spacetime generalisation of the Letaw-Frenet equations. The analytical solution for the radial LUA trajectories along with its past and future intercepts ${\cal C}$ with the past null infinity ${\cal J^-}$ and future null infinity ${\cal J^+}$ are presented. The Rindler horizons, in the presence of the black hole, are found to depend on both the magnitude of acceleration $|a|$ and the asymptotic initial data $h$, unlike in the flat Rindler spacetime case wherein they are only a function of the global translational shift $h$. The horizon features are discussed. The Rindler quadrant structure provides an alternate perspective to interpret the acceleration bounds, $|a| \leq |a|_b$ found earlier in arXiv:1901.04674. 

\end{abstract}

\section{Introduction} \label{Introduction}

The Rindler trajectory in the presence of a black hole reveals some curious features. In a general curved spacetime, a linearly uniformly accelerated (LUA) trajectory is defined as a trajectory described with a constant curvature equal to the magnitude of its $4$ - vector acceleration $|a|$, with vanishing torsion and hyper-torsion as defined in the sense of the Letaw-Frenet equations \cite{letaw, kolekar}. The LUA trajectories are then, by construction, locally Rindler at every point along the trajectory when viewed from a localized inertial frame. A class of such radial LUA trajectories were investigated in \cite{kajol} in the background of a Schwarzschild Black hole. A radially inward moving trajectory, starting from spatial infinity, approaches a closest distance $r_{min}$ from the black hole and then returns back to infinity. Interestingly, for finite asymptotic initial data $h$, an upper bound on the magnitude of acceleration ${|a|}_b$ was found to exist for the trajectory to turn back. For all values of acceleration $|a|>{|a|}_b$, the trajectory always falls into the black hole horizon. The distance of closest approach $r_{min}$ then has a lower bound $r_b> 2M$ greater than the Schwarzschild radius. An upper bound on acceleration and lower bound on distance of closest approach were shown to exist for all finite asymptotic initial data $h$. 

In flat spacetime, the Rindler trajectory is confined to the right hand wedge of the Minkowski spacetime, the Rindler quadrant, formed by the past null infinity, future null infinity, past horizon null surface $X=-T$ and future horizon null surface $X=T$, where $X$ and $T$ are Minkowski coordinates \cite{rindler}. The particular casual structure of the quadrant in the background Minkowski spacetime along-with the  time-like boost Killing trajectories leads to the celebrated result of Unruh that the Minkowski vacuum is thermal with a temperature proportional to the magnitude of acceleration of the Rindler trajectory \cite{davies,unruh}. In a general curved spacetime, it has been argued that one can construct, in principle, trajectories which are locally Rindler and associate a first law of thermodynamics with the corresponding local Rindler horizon by analysing the flow of matter flux through it \cite{local1, local2, local3, local4, local5}; although an explicit analytical solution for such trajectories in the literature is not known. It would then be interesting to obtain a solution for the Rindler trajectory in a curved spacetime and further analyse the corresponding Rindler quadrant structure in the presence of a black hole. The role of the flat spacetime uniformly accelerated hyperbolic trajectory being replaced by the LUA trajectory in the black hole spacetime, one can generally expect the accompanying  Rindler horizon and the Rindler quadrant too to be affected due to the background curvature. We consider the formal definition for the future horizon, namely the causal past of the intercept of the LUA trajectory at future null infinity. Similarly, the past horizon is the future of the intercept of the LUA trajectory with the past null infinity. As is evident from the acceleration bound, only trajectories with finite asymptotic initial data $h$ and acceleration $|a|<{|a|}_b$ have a turning point resulting in an interception with the future null infinity and hence, the Rindler quadrant exists only for the LUA trajectories satisfying the acceleration bound $|a| < {|a|}_b$.

The paper is presented as follows: we investigate the structure of the Rindler horizons and the corresponding Rindler quadrants for radial LUA trajectories in the Schwarzschild background satisfying the acceleration bound.
 In section \ref{acceleration bound}, we briefly summarize the earlier results in \cite{kajol} on acceleration bounds for radial LUA trajectories in the Schwarzschild spacetime and then compare them with uniformly accelerated stationary observers at fixed spatial Schwarzschild co-ordinates. The different possible parametrisations of the LUA trajectories in terms of $(|a|,h)$, $(|a|, r_{min})$ and $(|a|, {\cal C})$ are also discussed in section \ref{Alternate parametrization}. In section \ref{Solution}, we determine the explicit analytical solution for the radial LUA trajectory and find its asymptotic expansion in section \ref{intercept}. We then investigate the features associated with the Rindler horizons and discuss the structure of Rindler quadrant in section \ref{Quadrant}. In section \ref{Metric}, we present a form of the metric for the quadrants in the co-moving frame of LUA observer. The conclusions are presented in section \ref{discussion}. The signature of the metric is taken to be $(+,-,-,-)$.

\section{Acceleration bounds}\label{acceleration bound}

We briefly summarise below the results in \cite{kajol} pertaining to acceleration bounds for radial LUA trajectories in a Schwarzschild spacetime.  

A LUA trajectory in a curved spacetime is essentially locally Rindler, that is, locally it is a hyperbolic planar trajectory, at every point along the trajectory when viewed from a local inertial frame. The LUA trajectory, in addition to the constancy condition on the magnitude of acceleration, satisfies a further constraint of \textit{linearity} having vanishing torsion and hyper-torsion. In \cite{kolekar}, a construction based on the Letaw-Frenet equations and their corresponding geometrical scalar invariants was shown to lead to such a covariant definition of the linear uniformly accelerated (LUA) trajectory satisfying the following constraint equation
\begin{eqnarray}
w^{i}-{|a|}^{2}u^{i}=0
\label{linearity}
\end{eqnarray}
where $w^{i}=u^{j}{\nabla}_{j}a^{i}$ and $a^i$ and $u^i$ are the acceleration and velocity four vectors respectively. The solution $x^{i}(\tau)$ consistent with the above constraint equation and a constant $|a|$ in a given background curved spacetime is the trajectory of the linear uniformly accelerated (LUA) observer vis-a-vis the generalised Rindler trajectory.

In \cite{kajol}, we analysed the radial LUA trajectories in a spherically symmetric general background metric of the form
\begin{eqnarray}
{ds}^{2} &=& f(r) \, {dt}^{2}-{f(r)}^{-1}{dr}^{2}-r^{2}\,{d\theta}^{2}-r^{2}{\sin}^{2}\theta \, {d\phi}^{2}.
\label{general metric}
\end{eqnarray}
The solution for the radial LUA trajectory consistent with Eq.(\ref{linearity}) was found in terms of it's four velocity as,
\begin{eqnarray}
u^{0} &=& \dfrac{dt}{d\tau}={f(r)}^{-1}\left(|a|r+h\right)\label{temporal gen velocity}\\
u^{1} &=& \dfrac{dr}{d\tau}=\pm\sqrt{\left(|a|r+h\right)^2-f(r)}\label{radial gen velocity}\\
u^{2} &=& u^{3} \; =0.
\end{eqnarray}
where, $|a|$ is the constant magnitude of acceleration and $h$ specifies the initial data at spatial infinity which accounts for the non linear shift in the trajectory along the radial direction. Here the spacetime considered is a black hole for a class of smooth differentiable functions $f(r)$ such that  $f(r_{s})=0$ at some radius $r_{s}$ and the spacetime is asymptotically flat, $f(r)\to1$ at spatial infinity, $r\to\infty$. To obtain the explicit solution $x^i(\tau)$, one needs to provide the form of $f(r)$ along-with a suitable boundary or a initial condition for the trajectory. We consider radial LUA trajectories with finite asymptotic initial data $h$ starting from a large radial distance moving towards the black hole having a outward pointing acceleration $3$-vector. The trajectory approaches a closest radial distance $r=r_{min}$ to the black hole which is the turning point of the trajectory and then returns back to radial infinity. However, due to the curvature effects of the black hole, there is an upper bound $|a|_b$ on the magnitude of acceleration for such a turning point $r_{min}$ to exist. The trajectory having acceleration greater than the bound value, $|a| > |a|_b$ does not have a turning point and  must fall into the horizon at $r_s$. Since the metric being considered is static with the Killing vector $\mathbf{\Xi} = \partial_t$, we choose $t=0$ when $r=r_{min}$ as our boundary condition. For the Schwarzschild metric
\begin{eqnarray}
{ds}^{2} &=& \left(1-\frac{r_{s}}{r}\right) \, {dt}^{2}-{\left(1-\frac{r_{s}}{r}\right)}^{-1}{dr}^{2}-r^{2}\,{d\theta}^{2}-r^{2}{\sin}^{2}\theta \, {d\phi}^{2}
\label{Schwarzschild metric}
\end{eqnarray}
the radial LUA trajectory in Eqs.(\ref{temporal gen velocity}) and (\ref{radial gen velocity}) can be written as
\begin{eqnarray}
\frac{dt}{dr} &=& \pm {\left(1-\frac{r_s}{r}\right)}^{-1} \frac{\left(|a| \, r+h\right) \,\sqrt{r}}{|a| \, \sqrt{(r-r_{min})(r-r_{max})(r-r_n)}}
\label{EOM-Schwarzschid}
\end{eqnarray}
where $r_{min}$, $r_{max}$ and $r_n$ are the roots of the cubic polynomial, $r\,{\left(|a| r+h\right)}^2 -r+r_s$ and are given as,
\begin{eqnarray}
r_{min} &=& \frac{2}{3\,|a|}\left[ \frac{\sqrt{3+h^2}}{2} \left(\cos(\xi/3)+\sqrt{3}\sin(\xi/3) \right)-h \, \right]\label{minimum}\\
r_{n} &=& \frac{-2}{3\,|a|}\left[ \sqrt{3+h^2} \, \cos(\xi/3)+h \,\right]\label{negative root}\\
r_{max} &=& \frac{2}{3\,|a|}\left[ \frac{\sqrt{3+h^2}}{2}\left( \cos(\xi/3)-\sqrt{3}\sin(\xi/3) \right)-h \, \right]\label{maximum}
\end{eqnarray}
where, $\xi=\tan^{-1}(B/A)$ with, $A=(27{|a|}^{4}r_{s}+18{|a|}^{3}h-2{|a|}^{3}h^{3})$ and $B=\sqrt{4 \, {|a|}^{6} \, (3+h^2)^{3}-(A)^{2}}$. These three roots of the cubic polynomial satisfy the following relations:
\begin{eqnarray}
r_{min}+r_{n}+r_{max} &=& - \, \frac{2h}{|a|}\label{Addition of roots}\\
r_{min} \, r_{max} + r_{min} \, r_{n} + r_{max} \, r_{n} &=& \frac{h^2 -1}{{|a|}^2}\label{relation 3}\\
r_{min} \, r_{n} \, r_{max} &=& - \, \frac{r_s}{{|a|}^2}\label{Multiplication of roots}
\end{eqnarray}

The root $r_{min}$ gives the turning point of the LUA trajectory initially moving towards the black hole starting from radius $r_i>r_{min}$ while $r_{max}$ gives the turning point of the trajectory moving away from black hole starting from $r_i<r_{max}$. The root $r_n$ being negative does not have any physical significance. Further, the roots $r_{min}$ and $r_{max}$ are positive real only for values of acceleration $|a| \leq {|a|}_{b}$, the bound value, while they become imaginary for $|a| > {|a|}_{b}$ indicating that a LUA trajectory violating the acceleration bound always falls into the black hole. The radius $r_{min}$ corresponding to the bound value of acceleration gives the lower bound on the distance of closest approach $r_{b}$ for the return trajectory. The expressions for the bound on acceleration and on the distance of closest approach are
\begin{eqnarray}
{|a|}_{b} &=& \frac{2\left(-9 \, h + h^3 + \sqrt{{(3+h^2)}^3} \right)}{27 \, r_s }\label{3.c- bound}\\
{r}_{b} &=& \frac{2}{3 {|a|}_b} \left(\frac{\sqrt{3+h^2}}{2}-h\right) \label{3.c-extremum at bound}
\end{eqnarray}
The value of the bound on the distance of closest approach $r_b$ is positive only for asymptotic initial data $h<1$. Hence, altogether, in the Schwarzschild spacetime for a radial LUA trajectory to turn back at $r_{min}$ it should have initial data $h<1$ and magnitude of acceleration $|a|<{|a|}_b$.

The lower bound on the distance of closest approach $r_b$ is found to be greater than the Schwarzschild radius $r_s$ for all finite boundary data $h < 1$. A comparison with the Rindler trajectory in flat spacetime further elucidates the interesting character of this result.

Consider the $h=0$ case wherein the LUA trajectory matches with the Rindler hyperbola in flat spacetime at asymptotic infinity, with $(t,r)=(0,0)$ being the bifurcation point of the Rindler horizon. By increasing $|a|$ all the way upto infinity, the turning point of the Rindler trajectory $r_{rindler} = 1/|a|$ can be brought arbitrary closer to $r = 0$ or to the Rindler horizon at $t=r$. In the present case, one has introduced a black hole centred at $r=0$. Here too, one would have expected in general, the turning point $r_{min}$ to approach the Schwarzschild radius $r_s$ for a continuous increase in the magnitude of acceleration. The lower bound $r_{b}$ is still inversely proportional to $|a|$ in Eq.(\ref{3.c-extremum at bound}). However, increasing the acceleration $|a|$ beyond the bound $ |a| \leq 1/(\sqrt{27} M)$, simply thrusts the trajectory into the black hole horizon on crossing the lower bound radius $r_{b} = 3M$.  

Having summarised the results in \cite{kajol}, we obtain the analytical solution to the LUA trajectories and describe the structure of the Rindler quadrant in the next section. Before proceeding, it is instructive to compare the LUA trajectories described above with the stationary trajectories at fixed spatial co-ordinates in the Schwarzschild spacetime and discuss the different possible parametrizations of the trajectory space.

\subsection{Comparison with the stationary observer}\label{Comparison with the stationary observer}
\begin{figure}[h!]
\centering
\includegraphics[width=8.5cm, height=6.5cm]{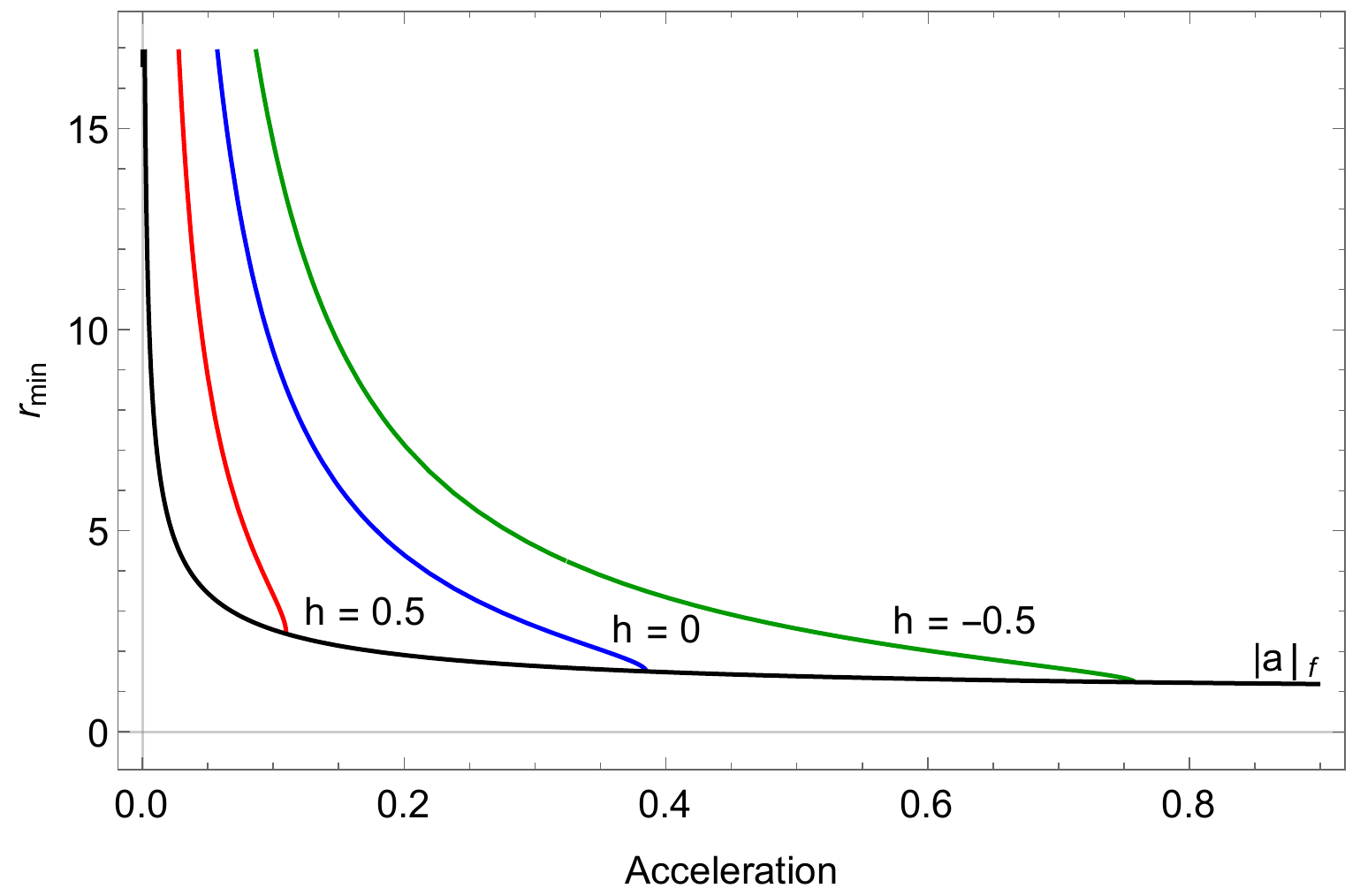}
\caption{The black curve represents the fixed radius $r$ versus $|a|_f$ for the stationary trajectory. The red, blue and green curves represent the distance of closest approach $r_{min}$ versus $|a|$ for the turning point LUA trajectories for $h=0.5$, $h=0$ and $h=-0.5$ respectively. The values of the bound on acceleration ${|a|}_b$ are $0.109927$, $0.3849$ and $0.758076$ for the red, blue and green curves respectively. Here $r_s=1$.} 
\label{comparison with stationary observer trajectory}
\end{figure}

The trajectory of a stationary observer in the Schwarzschild spacetime at the fixed spatial point ($r$, $\theta$, $\phi$) is also a LUA trajectory satisfying the linearity constraints in Eq.(\ref{linearity}), with the corresponding constant magnitude of acceleration dependent on the fixed radial coordinate $r$ and mass $M$ of the black hole through the relation
\begin{eqnarray}
{|a|}_f &=& {\left(1-\frac{r_s}{r}\right)}^{-1/2} \left(\frac{r_s}{2 \, r^2}\right)\label{acc of stationary observer}
\end{eqnarray}
A comparison of the fixed radius $r$ versus $|a|_f$ for the stationary trajectory and the distance of closest approach $r_{min}$ versus $|a|$ for the turning point LUA trajectories is shown in Figure \ref{comparison with stationary observer trajectory} for $r_s=1$.

For a non-stationary LUA trajectory, to have a turning point at $r_{min}$, its magnitude of constant acceleration at the turning point, when it is momentarily at rest, has to be greater than the magnitude of acceleration of the stationary observer at the same fixed radius $r=r_{min}$. In the contrary case, if the acceleration magnitude of the non-stationary LUA trajectory at any radius is less than the corresponding acceleration magnitude of the stationary trajectory at the same radius, then the inward \textit{gravity} of the black hole dominates the outward acceleration of the LUA trajectory and hence no turning point exists.  From Figure \ref{comparison with stationary observer trajectory}, it is evident from the $r_{min}$ versus $|a|$ curves that the non-stationary radial LUA trajectories always intersect the fixed radius $r$ versus $|a|_f$ curve for the stationary trajectory, at a finite value of $|a|_f$ with the corresponding fixed radius $r_{min}$ greater than the Schwarzschild radius $r_s$. Beyond the intersection point, $|a|$ for the non-stationary LUA trajectory is less than $|a|_f$  and hence no turning point exists. The finite value of  $|a|_f$ at the intersection points is then equal to the acceleration bound $|a|_b$ for a given asymptotic initial data $h$; and the corresponding fixed radius is the lower bound $r_b$ on the distance of closest approach $r_{min}$ which will always exist for all finite values of $h$. For the special case, when $|a| = |a|_b$, the acceleration of the non-stationary LUA trajectory at $r_{min}=r_b$ is equal to that of stationary observer at $r=r_b$. Hence, once the LUA trajectory reaches the distance of closest approach $r_b$, it remains stationary at radius $r_b$.

\subsection{Alternate Parametrisation}\label{Alternate parametrization}
The trajectory space can be equivalently parametrized by either of the pairs $(|a|,h)$, $(|a|, r_{min})$ or $(|a|,\mathcal{C})$ where $\mathcal{C}$ is the future (past) intercept of the trajectory with the future (past) null infinity as found in Eq.(\ref{Constant}). Here, we have used the algebraically transparent pair $(|a|,h)$ for our investigation, as the integration constant $h$ appears quite naturally in the expressions Eq.(\ref{temporal gen velocity} , \ref{radial gen velocity}) for the four velocity of the radial LUA trajectory in a general background metric of the Schwarzschild form in Eq.(\ref{general metric}). Nevertheless, $h$ has the physical interpretation of the parameter which accounts for the non linear shift in the trajectory along the radial direction as discussed in our earlier work \cite{kajol}. It was also shown, that for finite initial asymptotic data, that is keeping $h$ fixed, there exists an upper bound $|a|_b$ on the value of acceleration for the radially inward moving LUA trajectory to return back to infinity. The range of acceleration in this parametrization is $0 \leq |a| \leq |a|_b$ for a fixed $h$ which can have values $-\infty \leq h < 1$. For $h \rightarrow -\infty$, the upper bound value $|a|_b \rightarrow \infty$ while $r_{min} = r_b \rightarrow r_s$ and for $h \rightarrow 1$ the upper bound value $|a|_b \rightarrow 0$ while $r_{min} = r_b \rightarrow \infty$.

One could instead choose the geometrically transparent pair $(|a|, r_{min})$ to parametrize the trajectory. Then Figure \ref{comparison with stationary observer trajectory} suggests that keeping $r_{min}$ fixed there exist an infinite set of pairs $(|a|, h)$ satisfying Eq.(\ref{minimum}) for a particular value of $r_{min}$ and the lowest value of $|a|$ from that set would correspond to $|a|_b$ such that the radial LUA trajectory can return to infinity. This bound value $|a|_b$ would now be a lower bound value on acceleration and equal to the magnitude of acceleration of the uniformly accelerated stationary observer $|a|_f$ at $r=r_{min}$; essentially where the $r_{min}=$ constant horizontal line cuts the graph of $r$ versus $|a|_f$ and the chosen $r_{min}$ would be equal to $r_b$ in Eq.(\ref{3.c-extremum at bound}) for that value of $|a| = |a|_b$. The allowed range of acceleration in such parametrization is now instead ${|a|}_b \leq |a| \leq \infty$, for fixed $r_{min}$ which can take values $r_s \leq r_{min} \leq \infty$. For $r_{min} = r_b = r_s$, the lower bound value $|a|_b \rightarrow \infty$ and for $r_{min} = r_b \rightarrow \infty$, the lower bound value $|a|_b \rightarrow 0$.

Hence, it is possible to describe the bound on acceleration in either way, by keeping $h$ fixed or keeping $r_{min}$ fixed. 
In the former parametrization $(|a|,h)$, the bound appears as an upper bound while in the latter  parametrization $(|a|, r_{min})$ it appears as a lower bound. Both the descriptions are equivalent. 

To work with $(|a|, r_{min})$ parametrization, one needs to eliminate $h$ in favour of $|a|$ and $r_{min}$. This can be done by solving the cubic equation $r\,{\left(|a| r+h\right)}^2 -r+r_s=0$ with $r=r_{min}$ as,
\begin{eqnarray}
h(|a|,r_{min}) &=& \sqrt{1-\frac{r_s}{r}}-|a| r\label{h min}
\end{eqnarray}
Then using the relations in Eqs.(\ref{Addition of roots}), (\ref{Multiplication of roots}) and the expression for $h(|a|,r_{min})$, one can write the roots $r_{max}$ and $r_n$ completely in terms of $|a|$ and $r_{min}$ as,
\begin{eqnarray}
r_{max}(|a|,r_{min}) &=& \frac{r_{min}}{2}- \frac{1}{|a|} \sqrt{1-\frac{r_s}{r_{min}}} \nonumber \\
&& + \frac{1}{2|a|} \sqrt{4+{|a|}^2{r_{min}}^2-4 |a| r_{min} \sqrt{1-\frac{r_s}{r_{min}}}} \label{maximum in parametrisation 2}\\
r_{n}(|a|,r_{min}) &=& \frac{r_{min}}{2}-\frac{1}{|a|}\sqrt{1-\frac{r_s}{r_{min}}} \nonumber \\
&& -\frac{1}{2|a|}\sqrt{4+{|a|}^2{r_{min}}^2-4 |a| r_{min}\sqrt{1-\frac{r_s}{r_{min}}}}\label{negative root in parametrisation 2}
\end{eqnarray}
The above expressions for $r_{max}$ and $r_n$ together with $h$ in Eq.(\ref{h min}) then consistently satisfy the third relation given by Eq.(\ref{relation 3}). Also, one can further check that at the bound value of acceleration, $|a|=|a|_b$, $r_{max}$ is equal to $r_{min}$ as required. Thus it has been shown that the trajectory space can be re-parametrized by $(|a|, r_{min})$ simply by substituting $h=h(|a|, r_{min})$ with $r_{max}$ and $r_n$ as given by Eqs.(\ref{maximum in parametrisation 2}) and (\ref{negative root in parametrisation 2}) above, in the corresponding expressions with $(|a|, h)$ parametrization.

With the third possible parametrization $(|a|,\mathcal{C})$, choosing a fixed $\mathcal{C}$ value one can study all the trajectories corresponding to one Rindler quadrant in the Schwarzschild spacetime. However, to proceed with $(|a|,\mathcal{C})$, one needs to find the expression for $h$ as a function of $|a|$ and $\mathcal{C}$ by inverting Eq.(\ref{Constant}) which is possible in principle, but in practice it is algebraically quite complicated. 

In the present work, the algebraic expressions have been dealt with the earlier chosen $(|a|, h)$ parametrization in the following sections, while the graphs of geometric quantities such as the intercept ${\cal C}$ of the LUA trajectory with the null infinities, the bifurcation point of the Rindler horizons $r_{null}$ and the LUA trajectory curves are plotted in both the parametrizations $(|a|, h)$ and $(|a|, r_{min})$.

\section{Rindler Horizon in Schwarzschild spacetime}

As in the case of a Rindler trajectory in the flat spacetime, the future (past) Rindler horizon for the LUA trajectory in the  Schwarzschild spacetime is by definition, the causal past (future) of the future (past) intercept ${\cal C^+}$ (${\cal C^-}$) of the trajectory with future (past) null infinity ${\cal J^+}$ (${\cal J^-}$). The equations describing the future and past horizons are then given by the following outgoing and ingoing null geodesics having the same value for their corresponding intercepts respectively,
\begin{eqnarray}
r+ r_s \log\left|\frac{r}{r_s}-1\right|-t &=& \cal C^+ \nonumber \\
r+ r_s \log\left|\frac{r}{r_s}-1\right|+t &=& \cal C^- \label{future past Horizon}
\end{eqnarray}
Taking $r \rightarrow \infty$ limit in the equation of motion in Eq.(\ref{EOM-Schwarzschid}), one can note that the LUA trajectory asymptotes to null trajectories near radial infinity as expected and consistent with the fact that near spatial infinity, the trajectory tends to the usual hyperbolic Rindler trajectory in the flat spacetime. The leading terms in the series expansion near infinity of the equation of motion in Eq.(\ref{EOM-Schwarzschid}) are identical with those of the first integral of motion of null trajectories in the spacetime. Thus the intercepts ${\cal C^+}$ and ${\cal C^-}$ can be read-off formally as the constants in the asymptotic expansion of the solution $t(r)$ near $r \rightarrow \infty$. 

In the following section \ref{Solution}, we proceed to determine the explicit solution $t(r)$ for the LUA trajectory and then take its asymptotic expansion in section \ref{intercept}. Using these results, we then investigate the structure of the corresponding Rindler quadrant in section \ref{Quadrant}.

\subsection{Solution for a radial LUA trajectory}\label{Solution}

The explicit solution $t(r)$ for first integral of motion of the LUA trajectory in Eq.(\ref{EOM-Schwarzschid}) can be expressed in terms of the elliptic integrals as
\begin{eqnarray}
t(r) &=& \pm \; \sqrt{\frac{r(r-r_{min})(r-r_n)}{(r-r_{max})}}\nonumber\\
& & \pm  \, \frac{1}{|a|(r_s-r_{max})(r_s-r_{min})\sqrt{r_{min}(r_{max}-r_n)}} \nonumber\\
& & \quad \Big( \, 2 \, (h+|a| \, r_{max}) \, {(r_{max})}^2 \, (r_{min}-r_s) \; F\left(\Phi,M\right)\nonumber\\
& & \quad \; - |a| \, (r_{max}-r_s) \, (r_{min}-r_s) \, (r_{max}-r_n)\nonumber\\
& & \quad \; \quad \big( \; r_{min} \; E\left(\Phi,M\right)- \,(r_{min}-r_{max})\; F\left(\Phi,M\right) \, \big)\nonumber\\
& & \quad \; - 2 \, (h+|a| r_s) \, {(r_s)}^2 \, (r_{min}-r_{max}) \; \Pi \left(N_1,\Phi,M\right)\nonumber\\
 & & \quad \; + 2 \, |a|\, r_s \, (r_{max}-r_s) \, (r_{min}-r_s) \, (r_{min}-r_{max}) \, \Pi \left(N_2,\Phi,M\right) \Big)
\label{Solution with elliptic integrals}
\end{eqnarray}
with the $+$, $-$ signs referring to the outgoing and ingoing phases of the trajectory and the functions $\Phi$, $M$, $N_1$ and $N_2$ in terms of the radial co-ordinate $r$ are
\begin{align*}
\Phi &= {\sin}^{-1} \left( \sqrt{\frac{(r-r_{min})(r_{max}-r_n)}{(r-r_{max})(r_{min}-r_n)}} \right) & M & =\frac{r_{max}(r_{min}-r_n)}{r_{min}(r_{max}-r_n)}\\
N_1 &= \frac{(r_{max}-r_s)(r_{min}-r_n)}{(r_{max}-r_n)(r_{min}-r_s)} & N_2 &=\frac{(r_{min}-r_n)}{(r_{max}-r_n)}
\end{align*}
The incomplete elliptic integrals of the first, second and third kind are defined as
\begin{eqnarray}
F(\phi,m) &=& \int_0^\phi (1-m \sin^2\theta)^{-1/2} \, d\theta\\
E(\phi,m) &=& \int_0^\phi (1-m \sin^2\theta)^{1/2} \, d\theta\\
\Pi (n,\phi,m) &=& \int_0^\phi (1-n \sin^2\theta)^{-1} (1-m \sin^2\theta)^{-1/2} \, d\theta
\end{eqnarray}
Since the metric is independent of the time co-ordinate $t$, the solution is invariant under a time translation apart from an overall constant. We choose $t =0 $ when the trajectory is at its turning point $r = r_{min}$. This fixes the overall constant of integration and owing to symmetry about the $t = 0$ axis, we then have ${\cal C^+} = {\cal C^-}$. 

\begin{figure}[h!]
\begin{subfigure}{.5\textwidth}
\includegraphics[width=7cm,height=5cm]{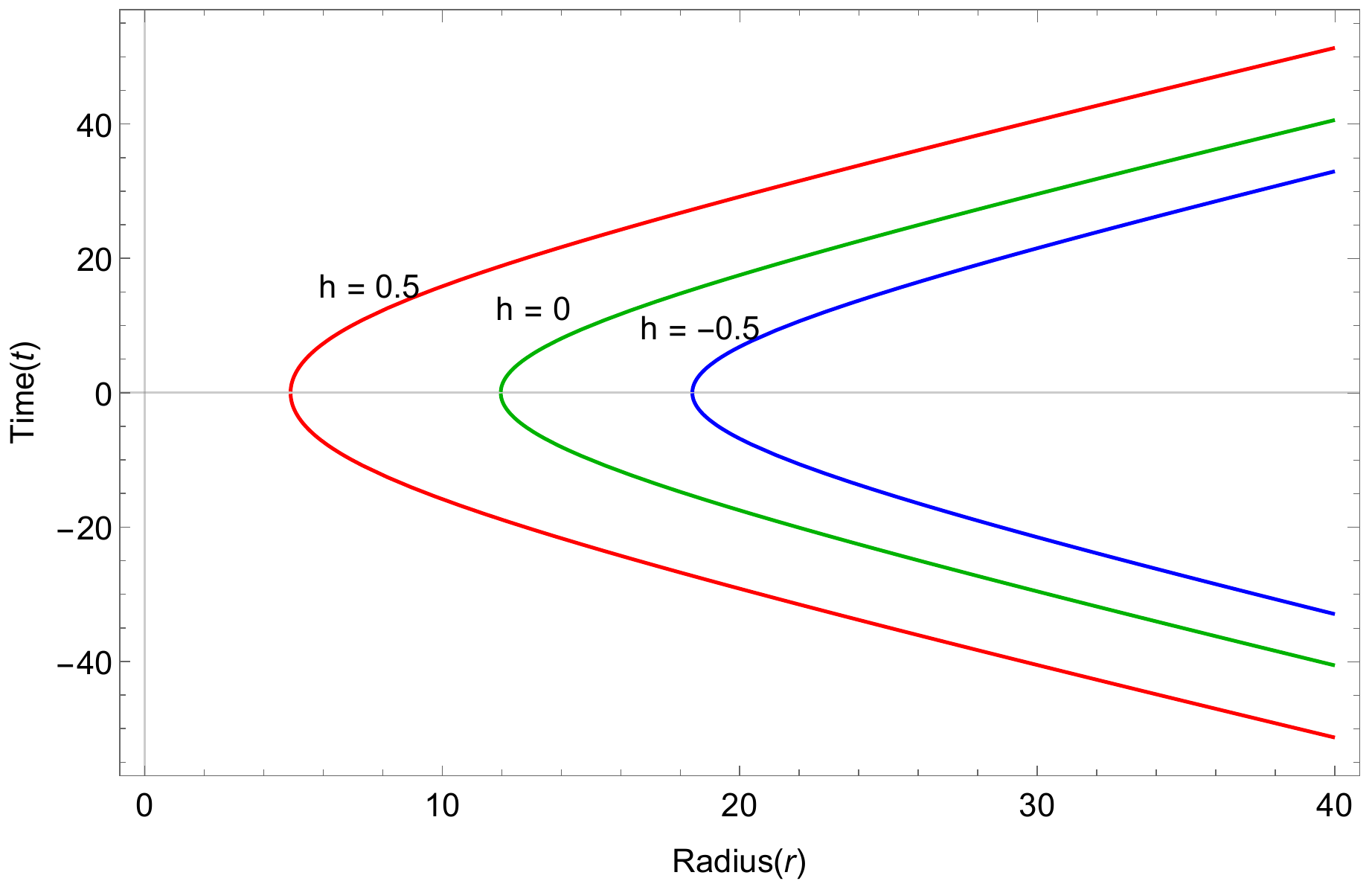}
\caption{$|a|= 0.08$}
\label{Trajectories with a=0.08}
\end{subfigure}
\begin{subfigure}{.5\textwidth}
\includegraphics[width=7cm,height=5cm]{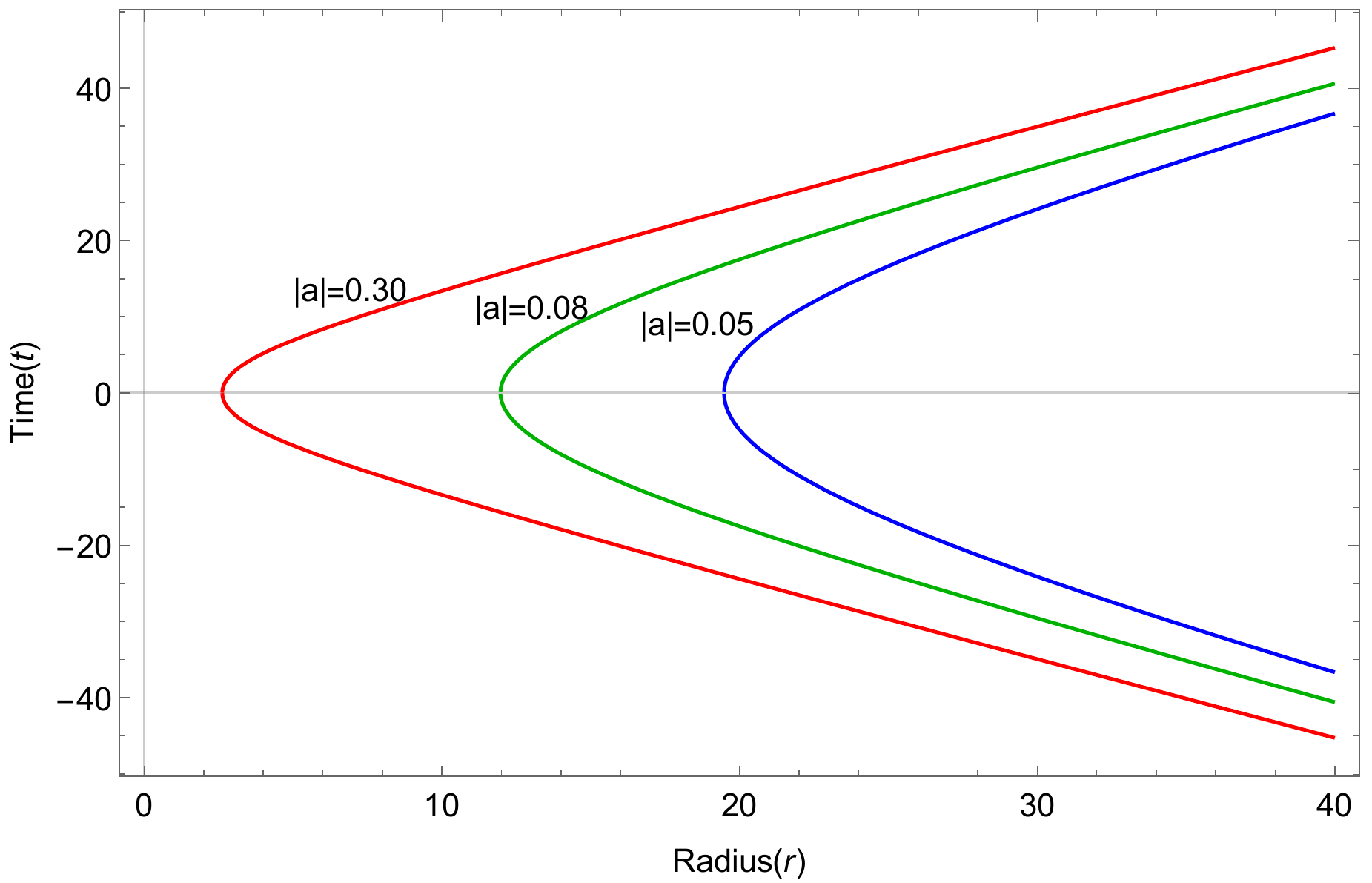}
\caption{$h=0$}
\label{Trajectories with h=0}
\end{subfigure}
\begin{subfigure}{.5\textwidth}
\includegraphics[width=7cm,height=5cm]{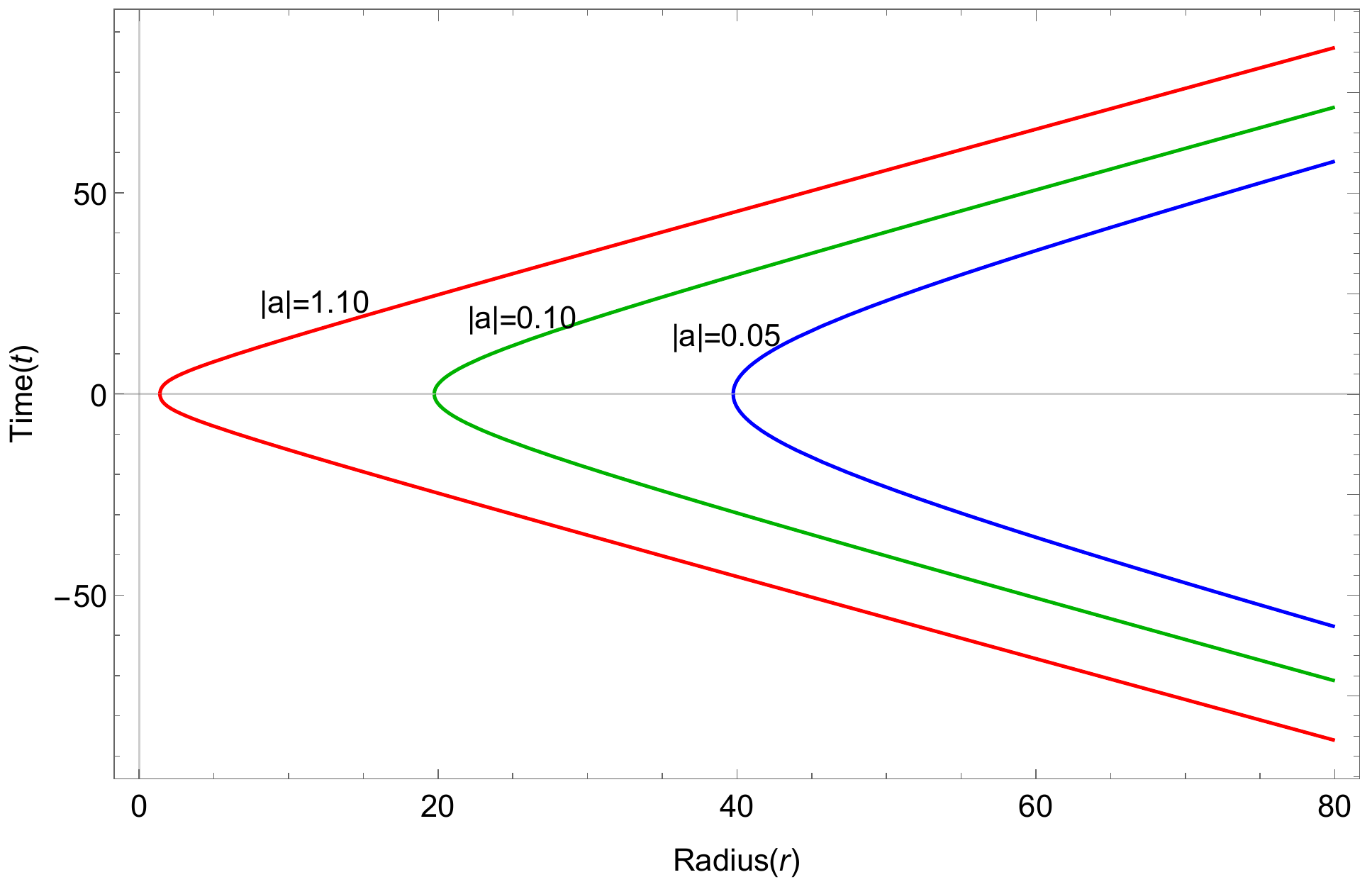}
\caption{$h=-1$}
\label{Trajectories with h=-1}
\end{subfigure}
\begin{subfigure}{.5\textwidth}
\includegraphics[width=7cm,height=5cm]{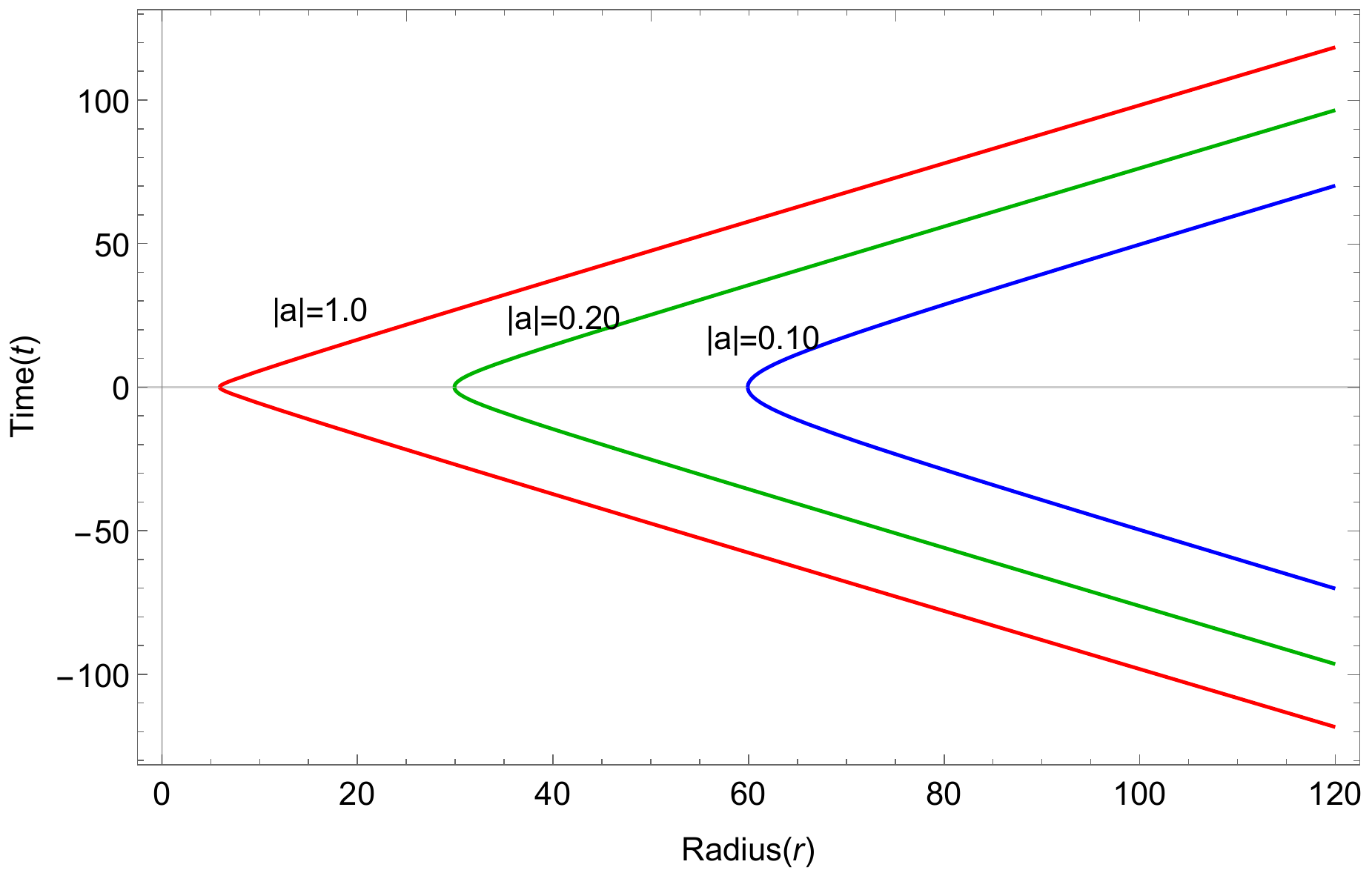}
\caption{$h=-5$}
\label{Trajectories with h=-5}
\end{subfigure}
\begin{subfigure}{.5\textwidth}
\includegraphics[width=7cm,height=5cm]{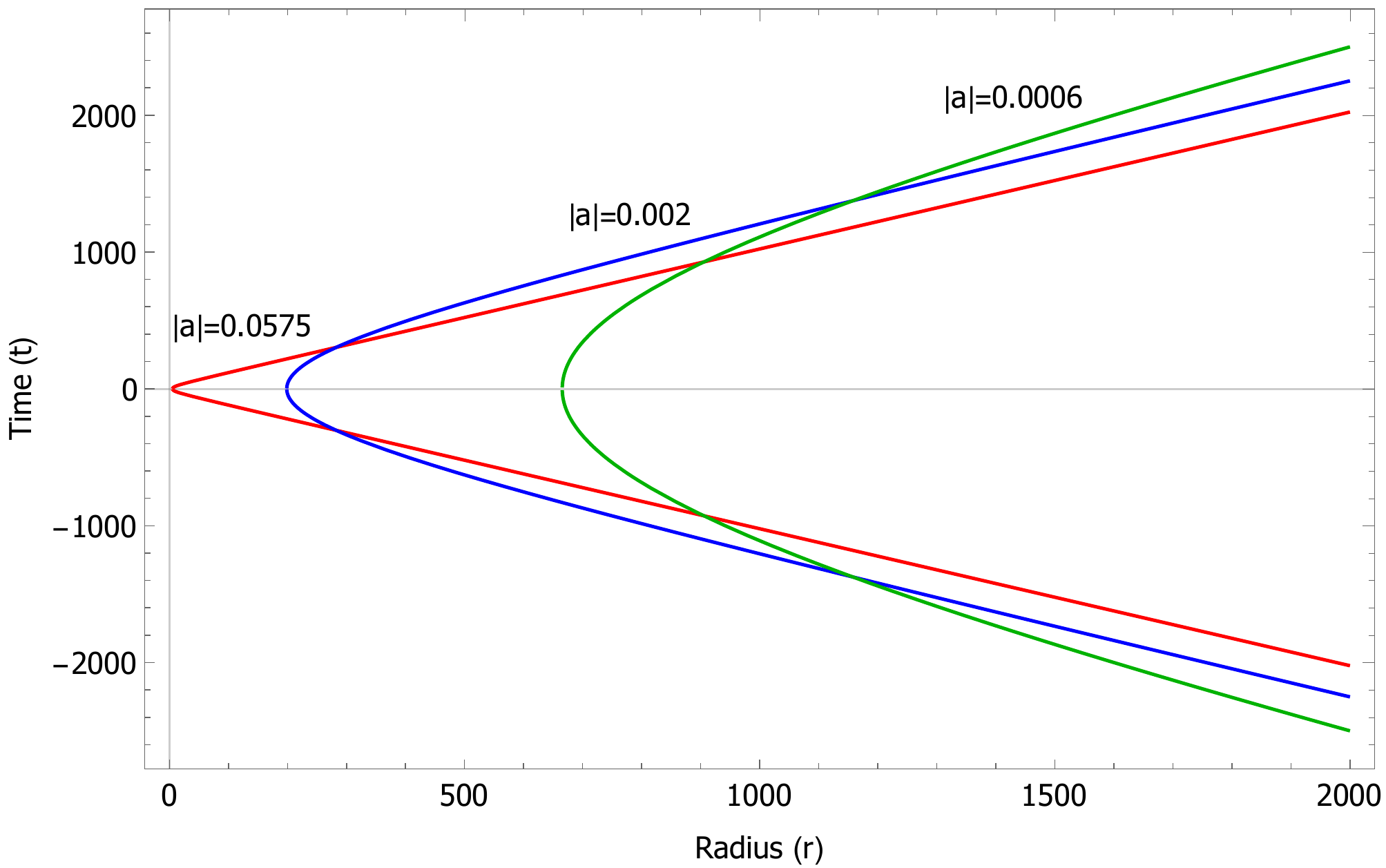}
\caption{$h= 0.6$}
\label{Trajectories with h=0.6}
\end{subfigure}
\begin{subfigure}{.5\textwidth}
\includegraphics[width=7cm,height=5cm]{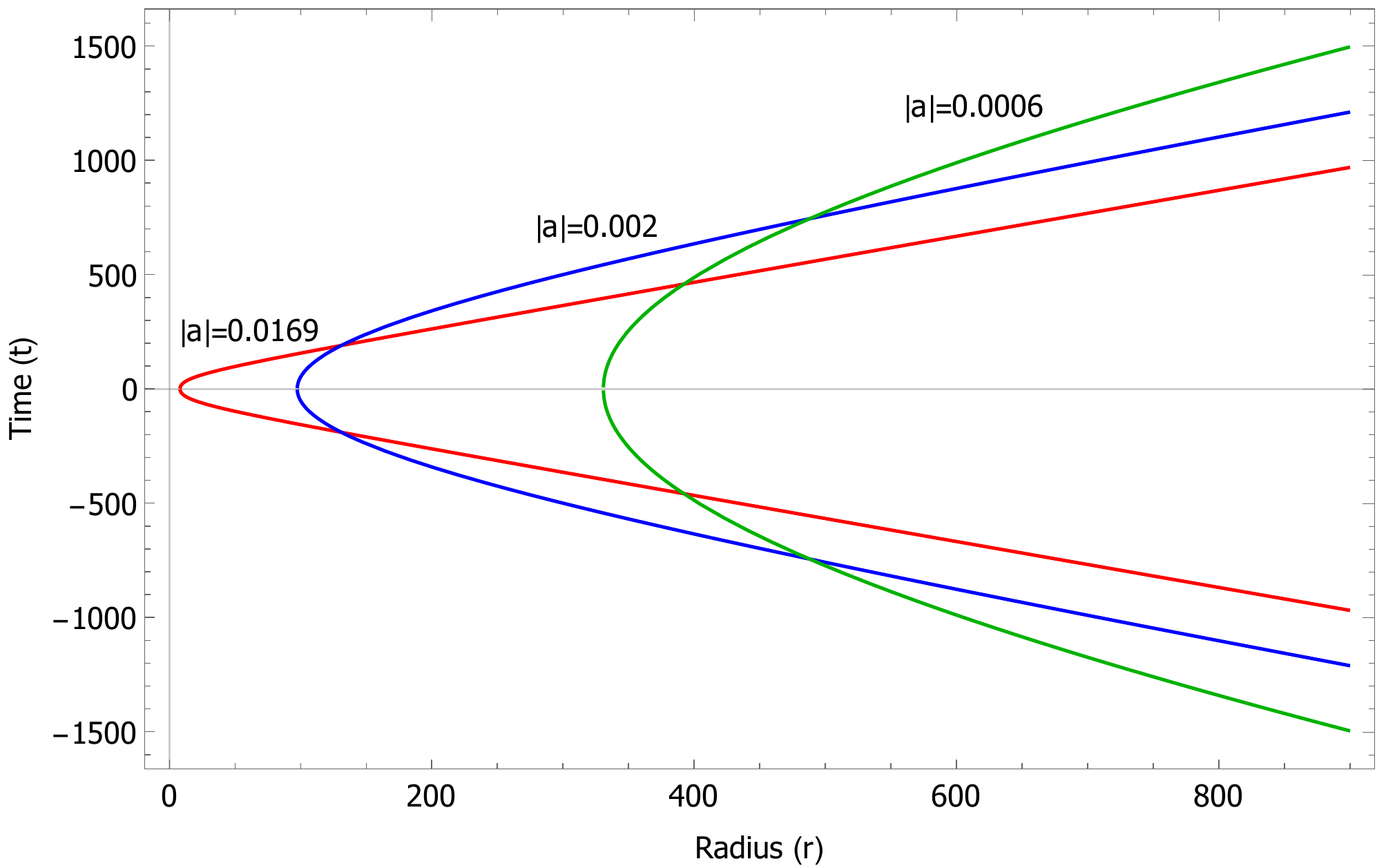}
\caption{$h=0.8$}
\label{Trajectories with h=0.8}
\end{subfigure}
\caption{LUA trajectories $t(r)$ for (a) $|a|=0.08$ and three different values of $h$ for which $|a|<{|a|}_b$ , (b) $h=0$ with ${|a|}_b=0.3849$ , (c) $h=-1$ with ${|a|}_b=1.1852$ , (d) $h=-5$ with ${|a|}_b=5.0490$ , (e) $h=0.6$ with ${|a|}_b=0.0722$ and (f) $h=0.8$ with ${|a|}_b=0.0190$.}
\label{LUA trajectories}
\end{figure}

To understand the broad nature of the trajectories, we have plotted them in the $(|a|,h)$ parametrisation for different values of acceleration $|a|$ and the asymptotic initial data $h$ classified under three cases, $h<0$, $h=0$ and $h>0$. In Figure \ref{Trajectories with a=0.08}, the plots are for three different values of $h$ with a fixed value of acceleration $|a|=0.08$, while satisfying the bound, whereas in Figures \ref{Trajectories with h=0} to \ref{Trajectories with h=0.8}, the plots are for a fixed value of $h$ with different values of acceleration $|a|$ less than the acceleration bound.

From the figures, one can observe that decreasing $h$ shifts the trajectory away from the black hole horizon, as expected, with the values of the asymptotic intercepts ${\cal C}$ being different for different values of $h$. However, unlike in the case of flat spacetime Rindler trajectories, in the Schwarzschild case for a fixed asymptotic initial data $h$, trajectories with different values of acceleration $|a|$, have different asymptotic intercepts implying the corresponding Rindler horizons to be different for each set $\{|a|, h \}$. Thus, in principle, we expect the formula for intercept ${\cal C}$ to be dependent on the asymptotic initial data $h$ as well as the magnitude of acceleration $|a|$ and the Schwarzschild radius $r_s$. 

For a fixed $h \leq 0$, increasing the acceleration $|a|$ accounts to increasing the value of the intercept $\cal C$. However, for a fixed $h>0$, 
increasing the value of acceleration $|a|$ increases the value of the intercept $\cal C$ till only a certain range of $|a|$ as shown in Figures \ref{Trajectories with h=0.6} and \ref{Trajectories with h=0.8}. As the distance of closest approach $r_{min}$ increases with decreasing acceleration, the trajectories with lower acceleration start to intersect the ones with higher acceleration and we then expect the intercept $\cal C$ to again increase with decreasing acceleration for a lower range of $|a|$. (The corresponding quantitative plots of $\cal C$ with respect to $|a|$ for a fixed $h>0$ are shown in Figure \ref{variation of C}.)

\begin{figure}[h!]
\begin{subfigure}{.5\textwidth}
\includegraphics[width=7cm,height=5cm]{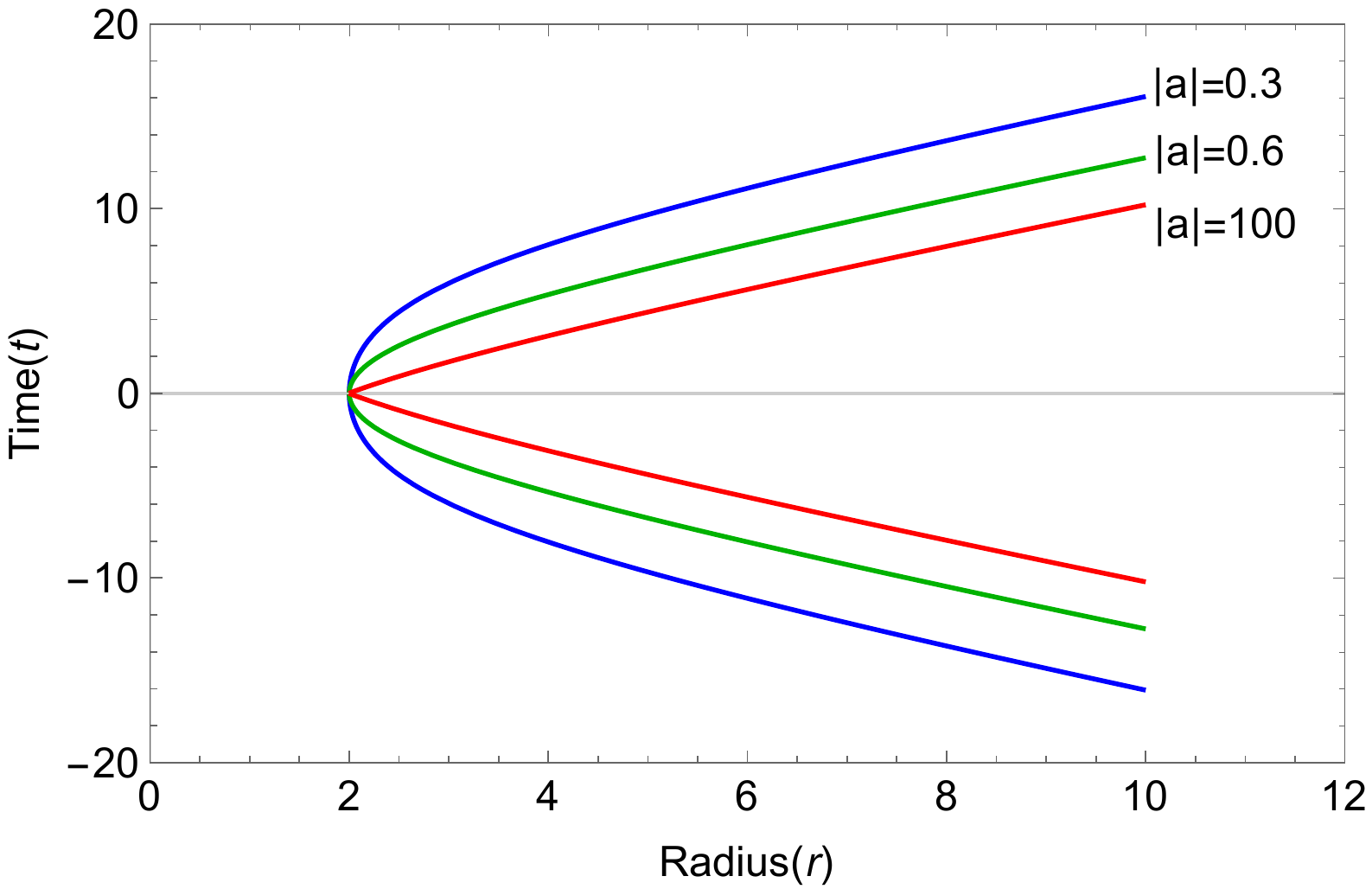}
\caption{$r_{min}=2$}
\label{Trajectories with rmin=2}
\end{subfigure}
\begin{subfigure}{.5\textwidth}
\includegraphics[width=7cm,height=5cm]{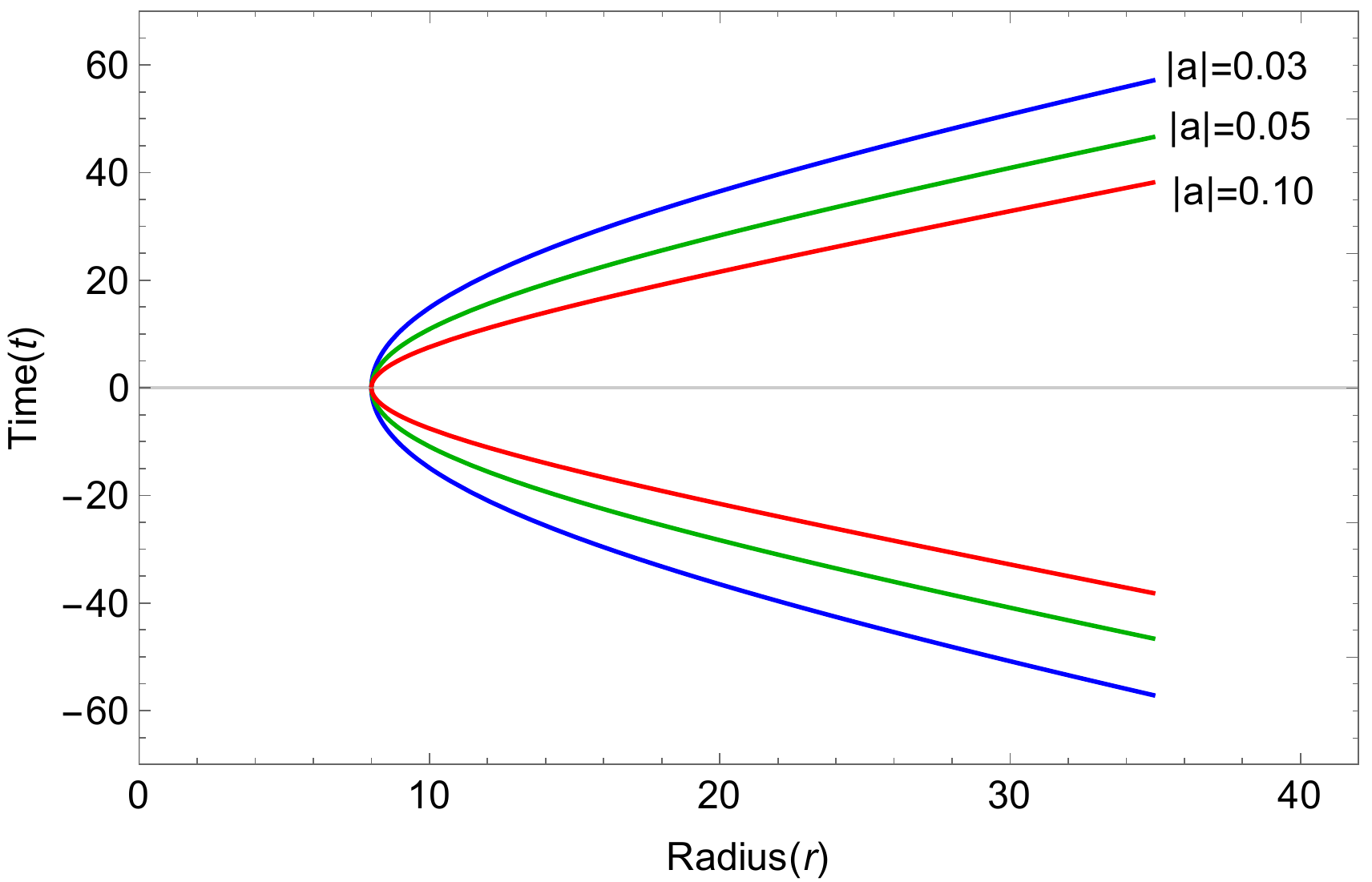}
\caption{$r_{min}=8$}
\label{Trajectories with rmin=8}
\end{subfigure}
\caption{LUA trajectories $t(r)$ for (a) $r_{min}=2$ with ${|a|}_b=0.1767$ and (b) $r_{min}=8$ with ${|a|}_b=0.0083$.}
\label{LUA trajectories with parametrization 2}
\end{figure}

In Figure \ref{LUA trajectories with parametrization 2}, we have also plotted the trajectories using the $(|a|, r_{min})$ parametrization and having same turning point $r_{min}$ with different magnitude of acceleration $|a|$ satisfying the lower bound $|a|>|a|_b$ , as explained in section \ref{Alternate parametrization}. The trajectories with different magnitude of acceleration $|a|$ have different asymptotic intercepts $\cal C$, which will be dependent on the parameters $|a|$ and $r_{min}$ as in the Rindler trajectories in the flat spacetime, but now the dependence being different than the flat case due to the effects of background curvature.

To explicitly obtain the expression of the intercept ${\cal C}$, we need the explicit asymptotic expansions of the elliptic integrals appearing in the  solution $t(r)$ in Eq.(\ref{Solution with elliptic integrals}). However, the elliptic integral of third kind $\Pi(N_2,\Phi,M)$ does not have a well defined expansion at radial infinity due to the ill-defined factor $(1 - n \sin^2\theta)^{-1}$ which diverges at $r \rightarrow \infty$. We circumvent the problem by re-deriving the solution $t(r)$ in a different form, namely as a series expansion which has a well defined asymptotic value at  $r\to\infty$. The procedure adopted is as follows. For a LUA trajectory starting from $r_i > r_{min}$, the relevant factors appearing in the denominator of the equation of motion in Eq.(\ref{EOM-Schwarzschid}) are expanded around $r_{max}$ and $r_s$ and the expression is written as a double summation as
\begin{eqnarray}
\frac{dt}{dr} &=& \pm \frac{1}{|a|} \, \sum_{i=0}^{\infty} \sum_{j=0}^{\infty} \binom{-1}{i} \binom{-1/2}{j} \frac{(-r_s)^i (-r_{max})^j}{(r)^{i+j}}  \frac{(|a|\, r+h)}{ \sqrt{(r-r_{min})(r-r_n)}} \label{EOM-Schwarzschild-Series}
\end{eqnarray}
where the notation $\binom{x}{y}$ refers to binomial coefficients. Integrating the above equation, we can write the solution $t(r)$ in terms of Hypergeometric functions as,
\begin{eqnarray}
t(r) &=& \pm \frac{2}{|a|} \; \sqrt{\frac{r_{min}\,(r-r_{min})}{r\,(r_{min}-r_n)}} \; \sum_{i=0}^{\infty} \; \sum_{j=0}^{\infty} \; G(i,j) \nonumber\\
 & & \Bigg(|a| \,(r_{min}-r_n) \; \mathbf{F_1} \left[\frac{1}{2},2-i-j,\frac{-1}{2},\frac{3}{2}; \frac{r-r_{min}}{r},\frac{-r_n \,(r-r_{min})}{r\,(r_{min}-r_n )}\right] \nonumber\\
 & & \quad + \, (h+|a|\, r_n) \; \mathbf{F_1} \left[\frac{1}{2},1-i-j,\frac{1}{2},\frac{3}{2};\frac{r-r_{min}}{r},\frac{-r_n \,(r-r_{min})}{r\,(r_{min}-r_n )}\right] \Bigg) \label{t solution}
\end{eqnarray}
where the coefficients $G(i,j)$ are defined to be
\begin{eqnarray}
G(i,j) &=& \binom{-1}{i} \binom{-1/2}{j} \; \left(\frac{-r_s}{r_{min}}\right)^i \left(\frac{-r_{max}}{r_{min}}\right)^j
\end{eqnarray}
The $\mathbf{F_1}$ is the Appell Hypergeometric Function expressed as a double series through
\begin{eqnarray}
\mathbf{F_1} \left( \alpha ,\beta ,{ \beta }^{'} ,\gamma ; x, y \right) &=& \sum_{p=0}^{\infty} \sum_{q=0}^{\infty} \frac{{( \alpha )}_{p+q} \; {(\beta)}_{p} \; {({\beta}^{'})}_{q} }{{( \gamma )}_{p+q} \; p\,! \; q\,!} \; {x}^{p} \; {y}^{q} \label{Appell}
\end{eqnarray}
which converges for $|x|<1$ and $|y|<1$ and for $\gamma$ not having vanishing or negative integer values. It is straightforward to verify that each term of the double series solution, Eq.(\ref{t solution}), satisfies the earlier chosen boundary condition $t(r_{min})=0$. 
At $r\to\infty$, the argument $(r-r_{min})/r \to 1$ and the Appell $\mathbf{F_{1}}$ function can be expressed in terms of a Hypergeometric function of a single variable using the relation,
\begin{eqnarray}
\mathbf{F_{1}} \left( \alpha ,\beta ,{ \beta }^{'} ,\gamma ; 1, y \right) &=& \frac{\Gamma(\gamma)\, \Gamma(\gamma-\alpha-\beta)}{\Gamma(\gamma-\alpha) \, \Gamma(\gamma-\beta)} \; \mathbf{F} \left( \alpha ,{ \beta }^{'} ,\gamma-\beta ; y \right)
\label{Appell to 2F1}
\end{eqnarray}
where the Hypergeometric $\mathbf{F}$ function is expressed as,
\begin{eqnarray}
\mathbf{F} \left( \alpha ,\beta ,\gamma\, ; y\right) &=& \sum_{p=0}^{\infty} \frac{{( \alpha )}_{p} \; {(\beta)}_{p} }{{( \gamma )}_{p} \; p\,!} \; {y}^{p}
\end{eqnarray}
For the two $\mathbf{F_1}$ functions in Eq.(\ref{t solution}), the value of the argument $(\gamma-\alpha-\beta)$ is $(i+j-1)$ and $(i+j)$ respectively. The gamma function in Eq.(\ref{Appell to 2F1}) is then ill-defined only for the cases $(i,j)=(0,0)$ and $(i,j)=(1,0),(0,1)$ where its argument becomes a negative integer and zero respectively. Hence, the relation in Eq.(\ref{Appell to 2F1}) is valid for all sets of values of $(i,j)$ except for $(0,0),(1,0),(0,1)$. For these separate three cases, we integrate the Eq.(\ref{EOM-Schwarzschild-Series}), express these three terms in the solution $t(r)$ in terms of elementary functions and then find their asymptotic expansions. The three terms $T_{ij}$ in the solution $t(r)$ can be simply written as,
\begin{eqnarray}
T_{00} &=& \sqrt{(r-r_{min})(r-r_n)}\nonumber\\
 & & + \frac{2h+|a|(r_{min}+r_n)}{2|a|} \log\left(\frac{\sqrt{r-r_n}+\sqrt{r-r_{min}}}{\sqrt{r-r_n}-\sqrt{r-r_{min}}}\right) \label{term 00}
\end{eqnarray}
\begin{eqnarray}
T_{01}+T_{10} &=& \frac{r_{max}+2 r_s}{|a|} \Bigg( |a| \log\left(\frac{\sqrt{r-r_{min}}+\sqrt{r-r_n}}{\sqrt{r_{min}-r_n}} \right)\nonumber\\
 & & \qquad \qquad \quad + \frac{h}{\sqrt{-r_n r_{min}}}\sin^{-1}\left(\sqrt{\frac{r_n(r_{min}-r)}{r(r_{min}-r_n )}}\,\right)\Bigg)
\end{eqnarray}
Here, again, the overall constant of integration is fixed by the condition that every term of the double series vanishes at $r=r_{min}$, that is $t(r_{min})=0$. Collecting all the terms together, the solution $t(r)$ for LUA trajectory is then written as
\begin{eqnarray}
t(r) &=& T_{00}+T_{01}+T_{10}+ \frac{2}{|a|} \; \sqrt{\frac{r_{min}\,(r-r_{min})}{r\,(r_{min}-r_n)}} \times \nonumber\\
& & \; \left( \sum_{i=1}^{\infty} \; \sum_{j=1}^{\infty} \; G(i,j) H(i+j) + \sum_{k=2}^{\infty} \; \left[G(0,k)+G(k,0)\right] H(k) \right)
\end{eqnarray}
where,
\begin{eqnarray}
H(k) &=& |a| \,(r_{min}-r_n) \; \mathbf{F_1} \left[\frac{1}{2},2-k,\frac{-1}{2},\frac{3}{2}; \frac{r-r_{min}}{r},\frac{-r_n \,(r-r_{min})}{r\,(r_{min}-r_n )}\right] \nonumber\\
 & & + \, (h+|a|\, r_n) \; \mathbf{F_1} \left[\frac{1}{2},1-k,\frac{1}{2},\frac{3}{2};\frac{r-r_{min}}{r},\frac{-r_n \,(r-r_{min})}{r\,(r_{min}-r_n )}\right]
\end{eqnarray}
We have thus found a series expansion form for the solution of the radial LUA trajectory in the Schwarzschild spacetime. 

\subsection{Asymptotic Solution}\label{intercept}
The asymptotic expansions of the terms $T_{00}$, $T_{01}$, $T_{10}$ being well defined near $r \rightarrow \infty$ can now be expressed as
\begin{eqnarray}
\left(T_{00}+T_{01}+T_{10}\right)\Bigg|_{r\to\infty} &=& r + r_s \log\left(\frac{4 \, r}{r_{min}-r_n}\right)  - \frac{r_{min}+r_n}{2}\nonumber\\
& &  + \, \frac{h \, (r_{max}+2 \, r_s)}{|a| \, \sqrt{-r_n \, r_{min}}} \; \sin^{-1}\left(\sqrt{\frac{-r_n}{r_{min}-r_n}}\,\right)
\label{expansion of first three terms}
\end{eqnarray}
Using Eq.(\ref{Addition of roots}), the third term in above expression can be re-expressed as $(h/|a|)+(r_{max}/2)$. Further, using Eqs.(\ref{Appell to 2F1}) and (\ref{expansion of first three terms}), the asymptotic solution for LUA trajectory can be written in the form, $t(r) = \pm \left( r + r_s \, \log\,(r/r_s) + \cal C \right)$ which matches the asymptotic form of the null trajectories representing the future and past Rindler horizons in Eq.(\ref{future past Horizon}). The intercept $\cal C$ can now be read-off to be
\begin{eqnarray}
\cal C &=&   \frac{h}{|a|} + \frac{r_{max}}{2} +  \frac{h \, (r_{max}+2 \, r_s)}{|a| \, \sqrt{-r_n \, r_{min}}} \, \sin^{-1}\left(\sqrt{\frac{-r_n}{r_{min}-r_n}}\,\right) \nonumber\\
 & & + \, r_s \, \log\left(\frac{4\, r_s}{r_{min}-r_n}\right)  + \frac{\sqrt{\pi}}{|a|} \; \sqrt{\frac{r_{min}}{r_{min}-r_n}} \; \left( S_d + S_s \right)
\label{Constant}
\end{eqnarray}
where $S_d$ and $S_s$ are double and single summation series respectively expressed as,
\begin{eqnarray}
S_d &=& \sum_{i=1}^{\infty} \; \sum_{j=1}^{\infty} \; G(i,j) \; \left[\: |a|\,(r_{min}-r_n)\,H_1(i+j) + (h+|a|\,r_n)\,H_2(i+j) \:\right] \label{Sd}\\
S_s &=& \sum_{k=2}^{\infty} \; \left(\,G(k,0)+G(0,k) \, \right) \; \left[ \: |a|\,(r_{min}-r_n)\,H_1(k) + (h+|a|\,r_n)\,H_2(k) \:\right]
\end{eqnarray}
with the functions being defined as
\begin{eqnarray}
H_1(k) &=& \frac{\Gamma (k-1)}{\Gamma \left(k-\frac{1}{2}\right)} \; \mathbf{F} \left[\frac{1}{2},\frac{-1}{2},k-\frac{1}{2};\frac{-r_n}{r_{min}-r_n}\right]\nonumber\\
H_2(k) &=& \frac{\Gamma (k)}{\Gamma \left(k+\frac{1}{2}\right)} \; \mathbf{F} \left[\frac{1}{2},\frac{1}{2},k+\frac{1}{2};\frac{-r_n}{r_{min}-r_n}\right]\nonumber
\end{eqnarray}
For $\cal C$ to be a finite valued number, the series $S_d$ and $S_s$ need to be convergent for all allowed values of $h$ and $|a|$. Since the single summation $S_s$ is just a special case of the double summation series $S_d$, proving the convergence of the latter is sufficient to prove the convergence of the intercept $\cal C$. We prove the convergence of $S_d$ using the comparison test as follows. From the definition of Hypergeometric functions, we write the following inequalities,
\begin{eqnarray}
\mathbf{F} \left[\frac{1}{2},\frac{-1}{2},i+j-\frac{1}{2};x\right] &<& \mathbf{F} \left[\frac{1}{2},\frac{1}{2},\frac{3}{2};x\right] = \frac{{\sin}^{-1}\sqrt{x}}{\sqrt{x}} < \frac{\pi}{2}\nonumber\\
\mathbf{F} \left[\frac{1}{2},\frac{1}{2},i+j+\frac{1}{2};x\right] &<& \mathbf{F} \left[\frac{1}{2},\frac{1}{2},\frac{3}{2};x\right] = \frac{{\sin}^{-1}\sqrt{x}}{\sqrt{x}} < \frac{\pi}{2}\nonumber\\
 & & \qquad \qquad \quad \forall \; (i+j) \geq 2 \; \textrm{and} \; |x|<1 \label{relation1}
\end{eqnarray}
 Further, the gamma functions can be expressed in terms of the Pochhammer symbol as,
 \begin{equation}
\binom{-n}{k} =  \frac{(-1)^k}{k!} \frac{\Gamma(n+k)}{\Gamma(n)} = \frac{(-1)^k}{k!}(n)_k \label{relation2}
\end{equation}
Using the above relations in Eqs.(\ref{relation1}) and (\ref{relation2})  
we arrive at the following inequalities,
\begin{eqnarray}
G(i,j) \, H_1(i+j) & < &  \frac{- {(1)}_{i+j} \; {(1)}_{i} \; {(1/2)}_{j} }{{(-1/2)}_{i+j} \; i\,! \; j\,!} {\left(\frac{r_s}{r_{min}}\right)}^{i} \; {\left(\frac{r_{max}}{r_{min}}\right)}^{j}\label{inequality1}\\
G(i,j) \, H_2(i+j) & < &  \frac{{(1)}_{i+j} \; {(1)}_{i} \; {(1/2)}_{j} }{{(1/2)}_{i+j} \; i\,! \; j\,!} {\left(\frac{r_s}{r_{min}}\right)}^{i} \; {\left(\frac{r_{max}}{r_{min}}\right)}^{j} \label{inequality2}
\end{eqnarray}
It is straightforward to check that the double summation over indices $i$ and $j$ of the left hand side of Eq.(\ref{inequality1}) and Eq.(\ref{inequality2}) is lesser than that of the right hand side which can be written as an Appell function of the form given in Eq.(\ref{Appell}) and hence convergent since $(r_s / r_{min})$ and $(r_{max} / r_{min})$  are always less than $1$, except for the saturated bound case $|a| = |a|_b$ where $(r_{max} / r_{min}) = 1$. However, in the saturated bound case the trajectory asymptotes to the $r_{min}$ value and the value of the intercept is infinity as expected. Thus we have proved the double series $S_d$ to be convergent and the intercept $\cal C$ to be finite valued for $h<1$ and acceleration $|a|$ satisfying the bound value, $|a| < |a|_b$. The other two cases, when $\cal C$ diverges is when $h=1$ and $h= -\infty$ as can be checked by taking the respective limits in Eq.(\ref{Constant}). 

\begin{figure}[h!]
\begin{subfigure}{.5\textwidth}
\includegraphics[width=7.5cm,height=5cm]{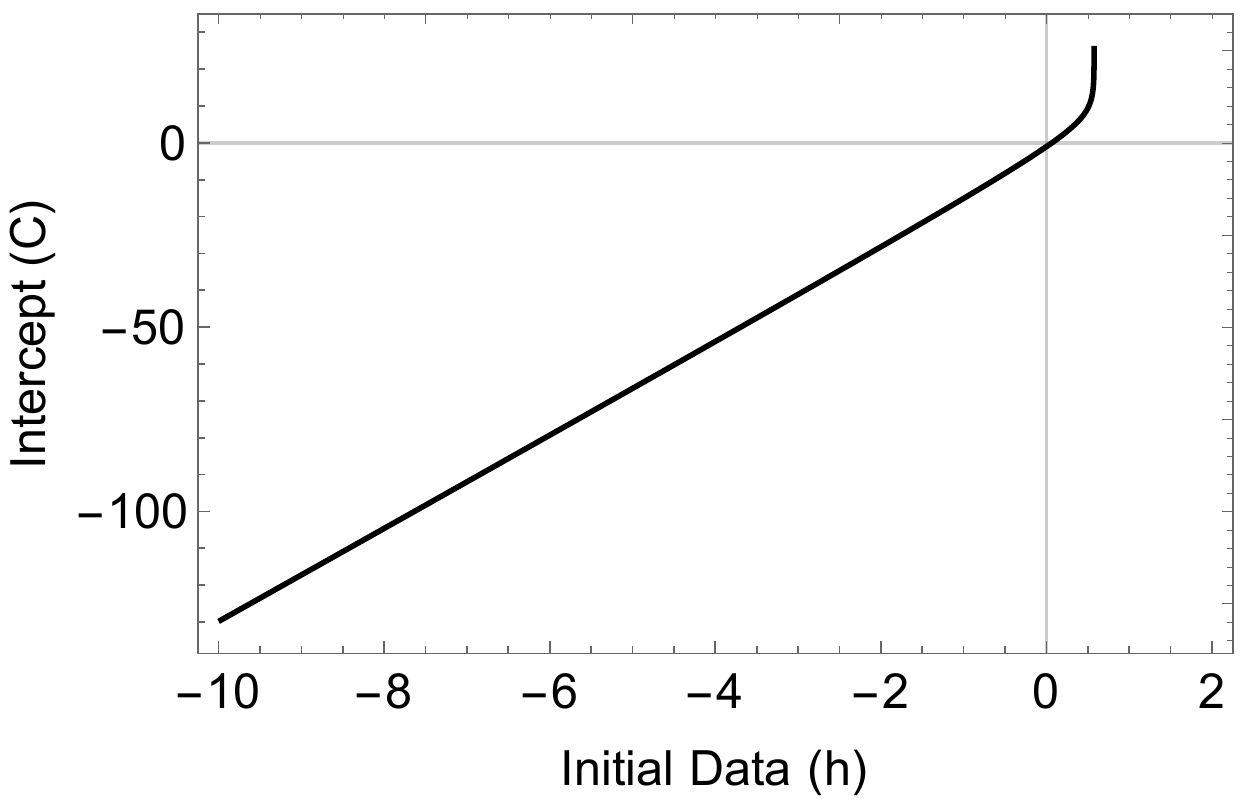} 
\caption{$|a|=0.08$}
\label{C vs initial data h with a=0.08}
\end{subfigure}
\begin{subfigure}{.5\textwidth}
\includegraphics[width=7.5cm,height=5cm]{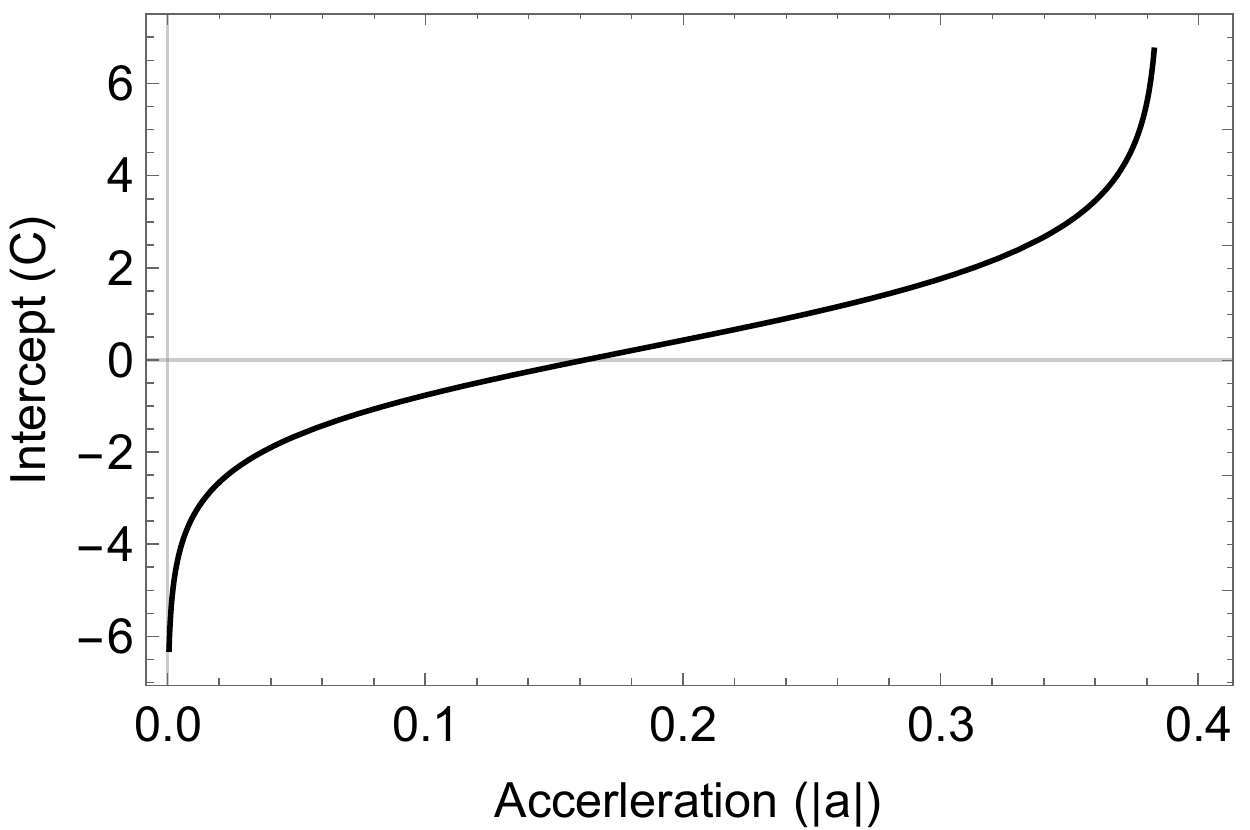} 
\caption{$h=0$}
\label{C vs acceleration with h=0}
\end{subfigure}
\begin{subfigure}{.5\textwidth}
\includegraphics[width=7.5cm,height=5cm]{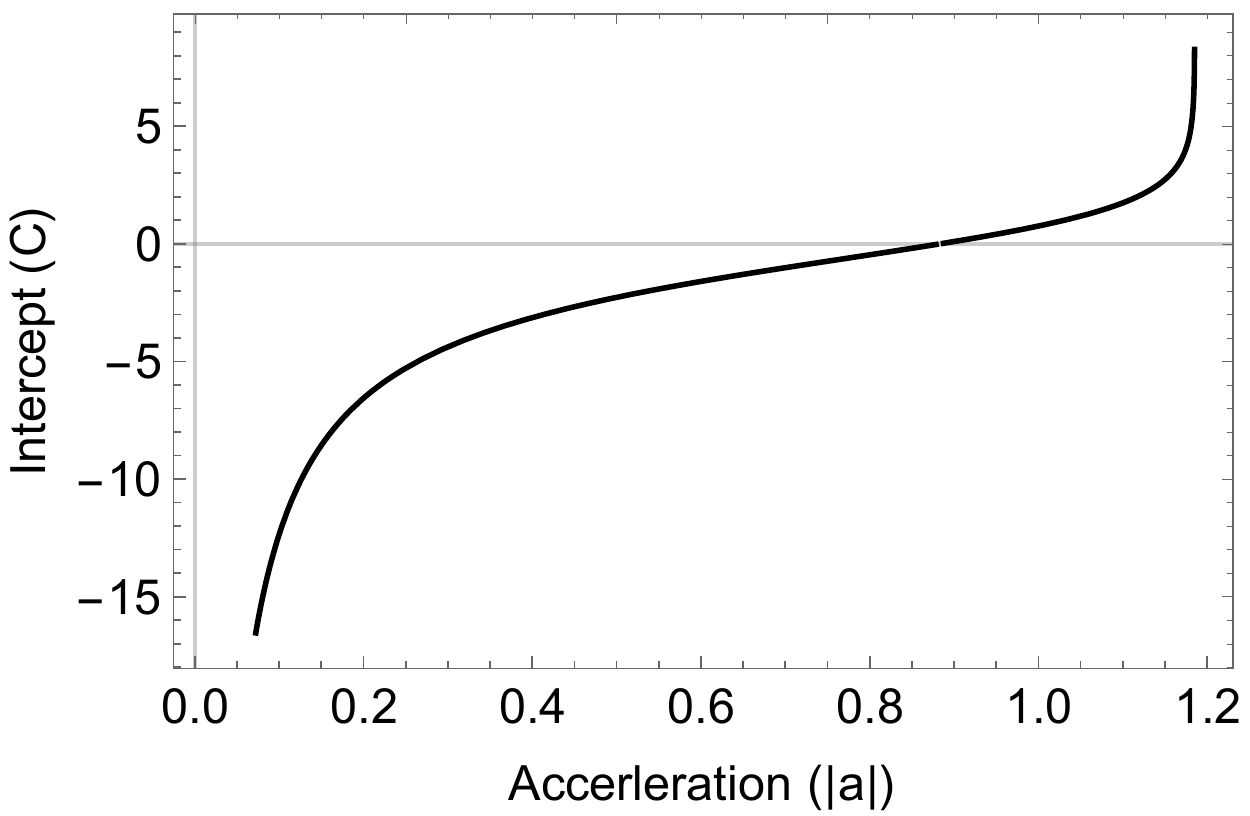} 
\caption{$h= -1$}
\label{C vs acceleration with h=-1}
\end{subfigure}
\begin{subfigure}{.5\textwidth}
\includegraphics[width=7.5cm,height=5cm]{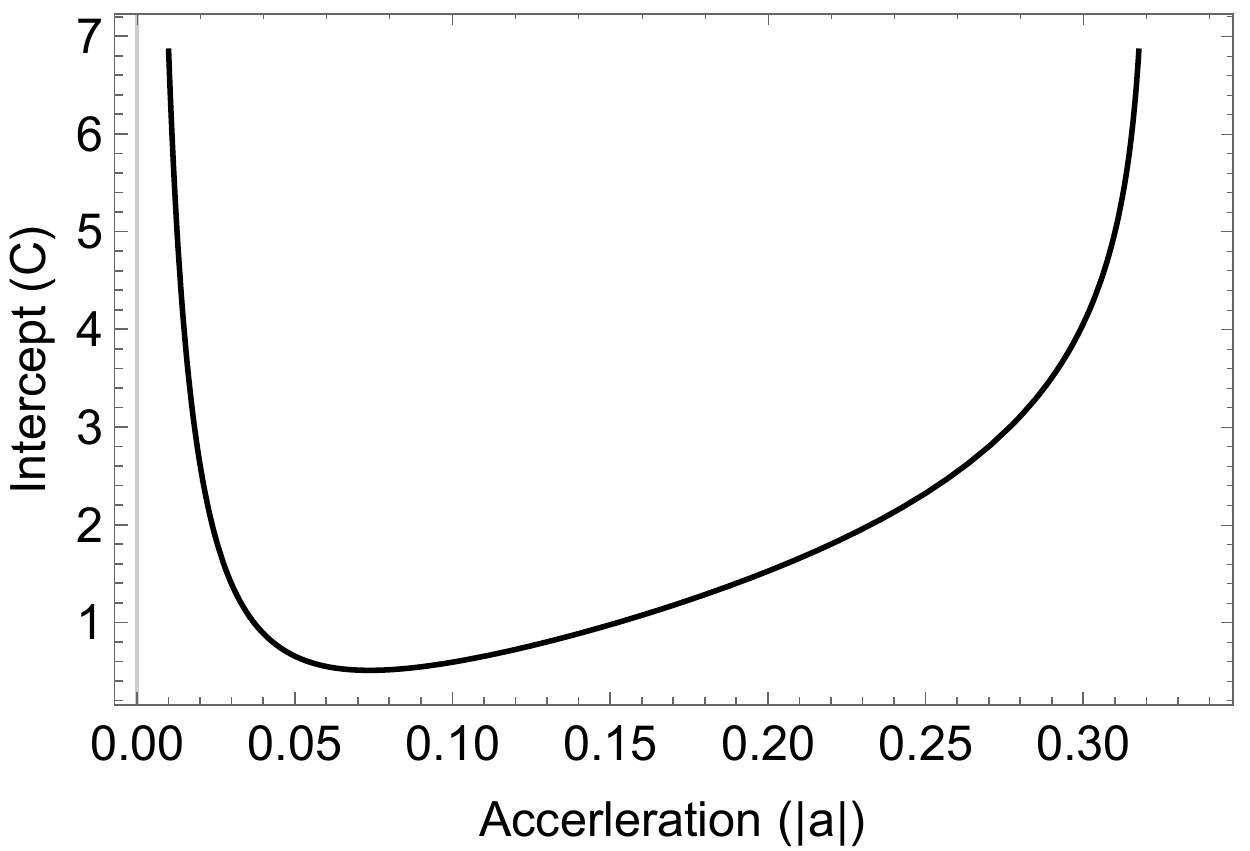} 
\caption{$h=0.1$}
\label{C vs acceleration with h=0.1}
\end{subfigure}
\caption{Variation of intercept $\cal C$ with initial data $h$ for acceleration $|a|=0.08$ and with acceleration $|a|$ for three different $h$ values, $h=0$, $h=-1$ and $h=0.1$. The bounds on the acceleration $|a|_b$ for $h=-1$, $h=0$ and $h=0.1$ are $1.1852$, $0.3849$ and $0.3202$ respectively.}
\label{variation of C}
\end{figure}

We have also evaluated the values of intercept $\cal C$ numerically by performing the single and double summation $S_s$ and $S_d$ in Mathematica by keeping terms upto 200 in each of the summations. The final result is found to be convergent in the order of $10^{-8}$ for the next 50 terms. The graphs of $\cal C$ against $h$ for a fixed value of acceleration $|a|=0.08$, and against $|a|$ for fixed value of asymptotic initial data $h$ for $h=-1$, $h=0$ and $h=0.1$ are shown in Figure \ref{variation of C}. The Schwarzschild radius is taken to be $r_s=1$ in all the cases. One can observe that for acceleration value close to the bound $|a|_b$, the value of intercept ${\cal C}$ approaches a large number as expected, since the intercept asymptotes to infinity as the acceleration $|a|$ approaches the bound value ${|a|}_b$.

\begin{figure}[h!]
\begin{subfigure}{.5\textwidth}
\includegraphics[width=7.5cm,height=5.5cm]{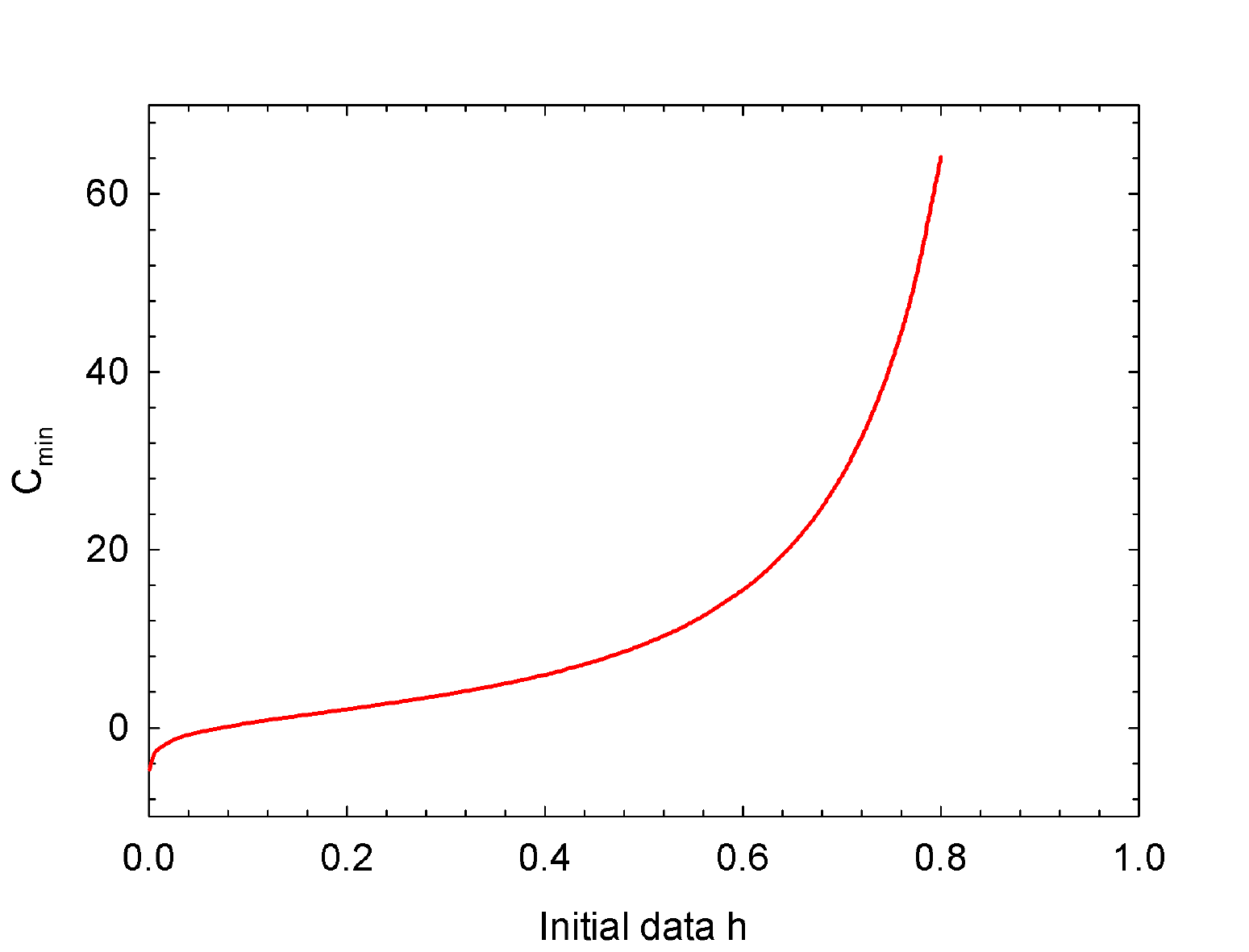} 
\caption{}
\label{Variation of C-min}
\end{subfigure}
\begin{subfigure}{.5\textwidth}
\includegraphics[width=7.5cm,height=5.5cm]{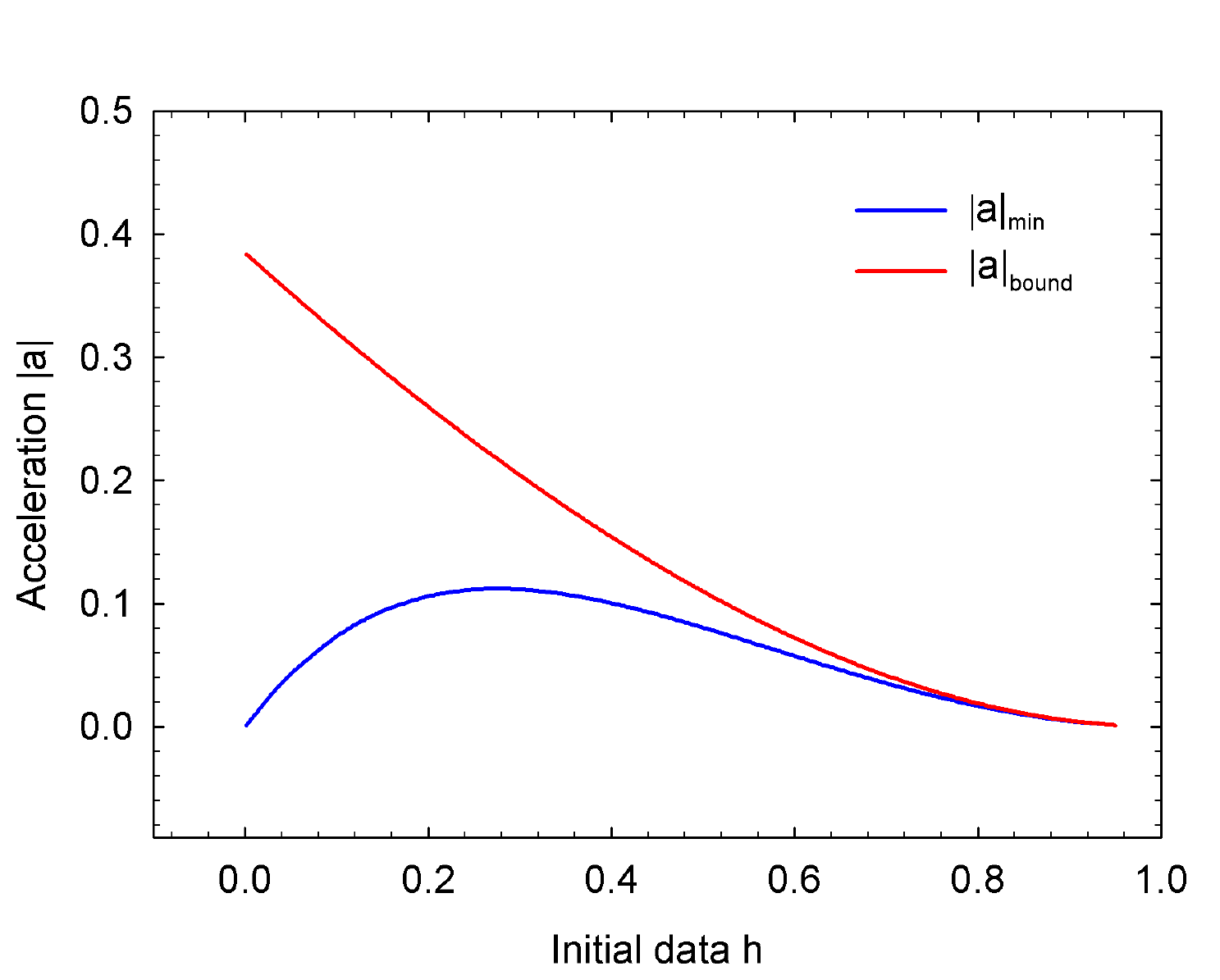} 
\caption{}
\label{Variation of a-min}
\end{subfigure}
\caption{Variation of ${\cal C}_{min}$ and ${|a|}_{min}$ with initial data value $h$ for $h>0$.}
\label{variation of C-min and a-min}
\end{figure}

For $h\leq 0$, the value of intercept $\cal C$ always increases monotonically with increasing value of acceleration $|a|$. While for the $0< h <1$ case, there exists a minimum value for $\cal C$, say ${\cal C}_{min}$ corresponding to a  ${|a|}_{min}$, such that all the trajectories having acceleration $|a| \neq {|a|}_{min}$ and $|a| < {|a|}_{b}$ have a broader shape than the one with acceleration ${|a|}_{min}$ at radial infinity, in the sense that a trajectory with acceleration ${|a|}_i<{|a|}_{min}$ will intersect all the trajectories having acceleration $|a|$ such that ${|a|}_i <|a| \leq {|a|}_{min}$. These observations are consistent with those in Figure \ref{LUA trajectories} obtained using the elliptic integrals solution. The values of ${\cal C}_{min}$ and ${|a|}_{min}$ are plotted in Figure \ref{variation of C-min and a-min} for particular values of $h$. 
From Figure \ref{Variation of a-min}, we can see that with increasing $h$ the range of acceleration ${|a|}_{min}<|a|<|a|_b$ decreases, that is, the number density of trajectories crossing the trajectory with acceleration ${|a|}_{min}$ decreases. 

These observations are in contrast with those for a Rindler trajectory in the flat spacetime. In the latter, for a fixed asymptotic initial data $h$, the trajectories with different $|a|$ do not intersect and are the integral curves of a vector field, namely the integral curves of the boost Killing vector constrained by common future and past horizons. In the Schwarzschild case, integral curves of the unique time-like Killing vector $\mathbf{\Xi} = \partial_t$ correspond to the stationary LUA trajectories at fixed spatial co-ordinates whereas the non-stationary LUA trajectories which form the main focus of the present paper, neither have a one to one correspondence to a Killing vector nor do they even correspond to integral curves of any time-like vector \textit{field}, for a fixed $h$.

The variation of intercept ${\cal C}$ is also studied with $(|a|, r_{min})$ parametrization. The plots are shown below in Figure \ref{variation of C parametrization 2} with $r_s=1$. Here $\cal C$ is a monotonically decreasing function of both $|a|$ and $r_{min}$. In Figure \ref{variation of C for a=0.5} the intercept ${\cal C}$ is plotted versus the turning point $r_{min}$ for a fixed value of acceleration $|a|=0.5$ which is a bound value of acceleration for $r_{min}=r_b=1.38$. Thus for $|a|=0.5$ as $r_{min}$ approaches $1.38$, the intercept $\cal C$ approaches infinity. The variation of intercept $\cal C$ with the magnitude of acceleration $|a|$ is shown in Figures \ref{variation of C for rmin=1.5}-\ref{variation of C for rmin=1.59} for three different values of $r_{min}$, $r_{min}=1.5$, $1.125$ and $1.59772$. For these three values of $r_{min}$ the lower bounds on the value of acceleration $|a|_b$ are $0.3849$, $1.1852$ and $0.3202$ respectively.
\begin{figure}[h!]
\begin{subfigure}{.5\textwidth}
\includegraphics[width=7cm,height=5cm]{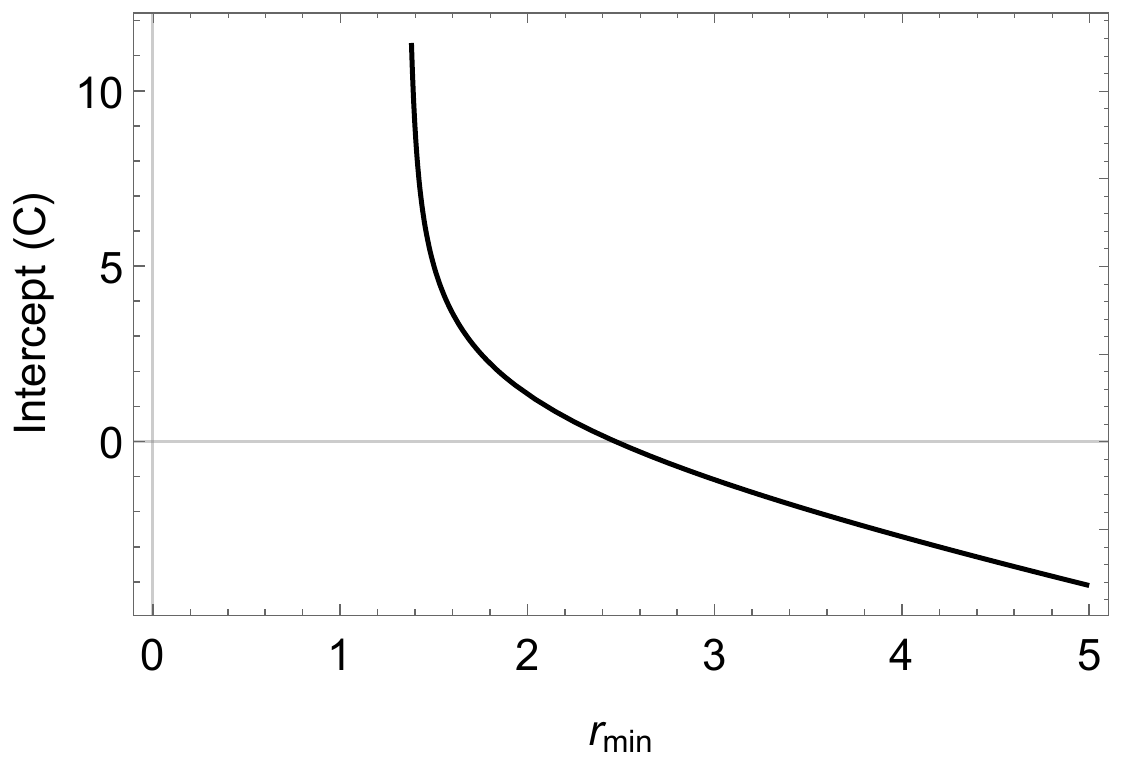}
\caption{$|a|= 0.5$}
\label{variation of C for a=0.5}
\end{subfigure}
\begin{subfigure}{.5\textwidth}
\includegraphics[width=7cm,height=5cm]{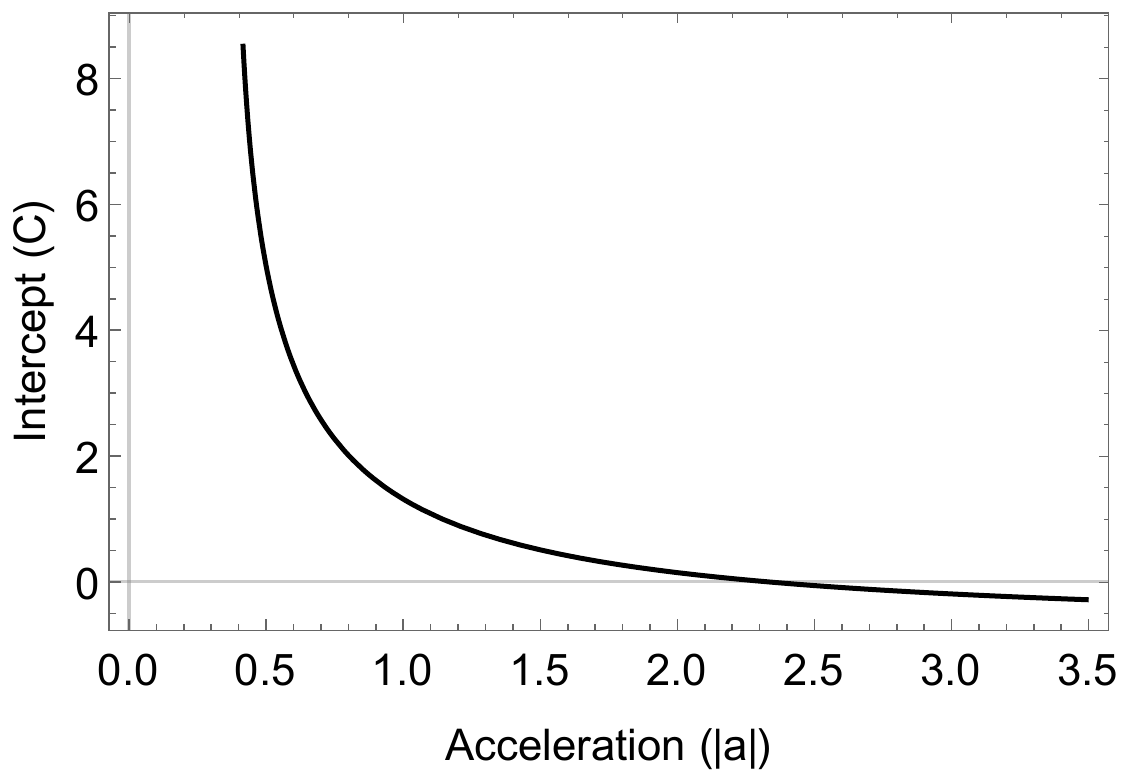}
\caption{$r_{min}=1.5$}
\label{variation of C for rmin=1.5}
\end{subfigure}
\begin{subfigure}{.5\textwidth}
\includegraphics[width=7cm,height=5cm]{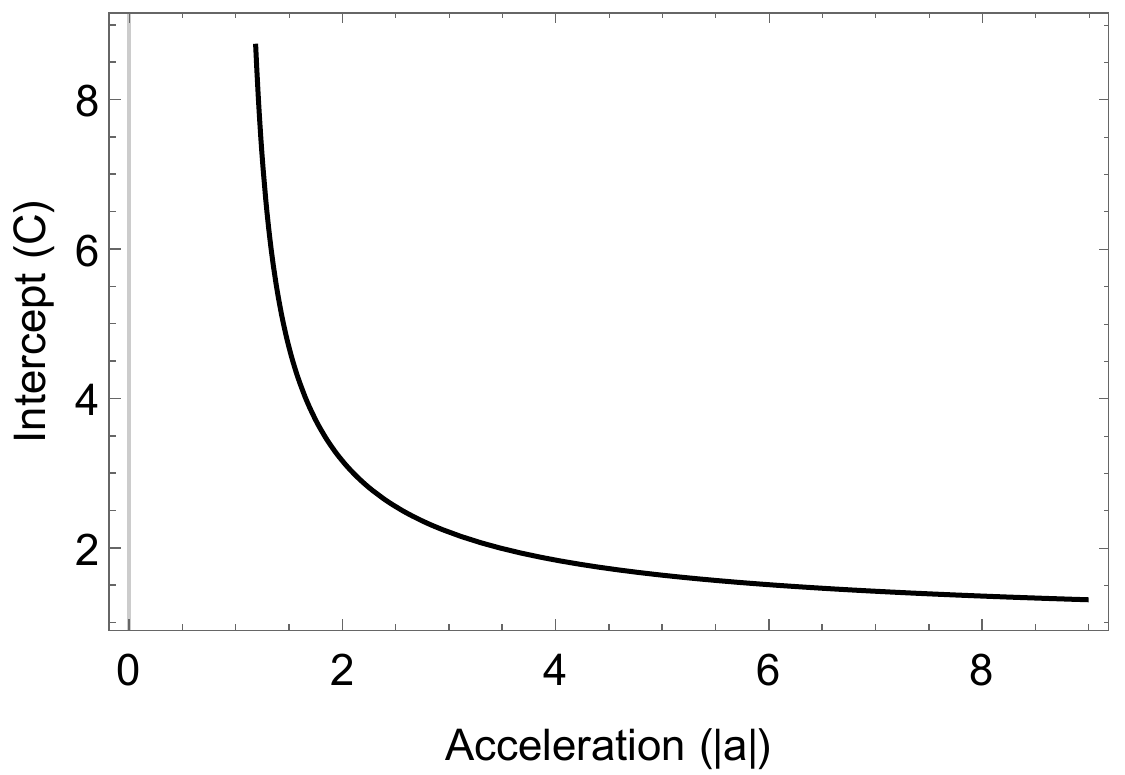}
\caption{$r_{min}=1.125$}
\label{variation of C for rmin=1.125}
\end{subfigure}
\begin{subfigure}{.5\textwidth}
\includegraphics[width=7cm,height=5cm]{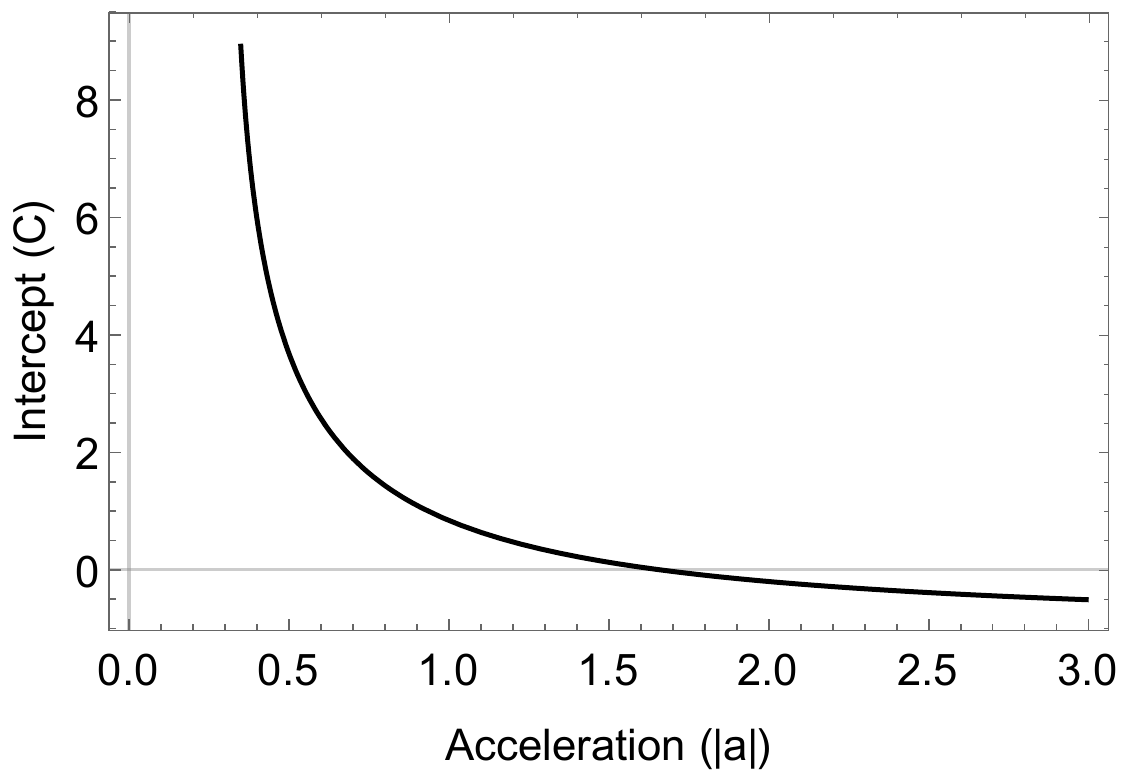}
\caption{$r_{min}=1.59$}
\label{variation of C for rmin=1.59}
\end{subfigure}
\caption{Variation of intercept $\mathcal{C}$ with $r_{min}$ for $|a|=0.5$ and with acceleration for three different $r_{min}$ values, $r_{min}=1.5$, $r_{min}=1.125$, and $r_{min}=1.598$.}
\label{variation of C parametrization 2}
\end{figure}

\subsection{Rindler Quadrant}\label{Quadrant}

The Rindler quadrant for the radial LUA trajectories in the Schwarzschild spacetime is then the union of the past and future null infinity with the past and future horizons in the Penrose diagram. The intersection points of these are the spatial infinity ${ i^0}$, the past and future intercepts $\cal C^+$ and $\cal C^-$ and the bifurcation point of the past and future horizons. The intercepts $\cal C^+$ and $\cal C^-$ were determined in the previous section \ref{intercept}. 

To determine the bifurcation point, we solve for the intersection point of the null geodesics corresponding to the future and past Rindler horizons in Eq.(\ref{future past Horizon}), with $\cal C$ given by Eq.(\ref{Constant}). We get
\begin{eqnarray}
r_{null} &=& r_s \left( 1+ W\left[e^{-1-\frac{\cal C}{r_s}}\right] \right)
\end{eqnarray}
where $W[x]$ is the productlog function. Thus, unlike in the case of the Rindler quadrant in the flat spacetime, for the Schwarzschild case, the bifurcation point $r_{null}$ is dependent on the asymptotic initial data $h$ as well as the acceleration magnitude $|a|$ and the Schwarzschild radius $r_s$.

\begin{figure}[h!]
\begin{subfigure}{.5\textwidth}
\includegraphics[width=7.5cm, height=5cm]{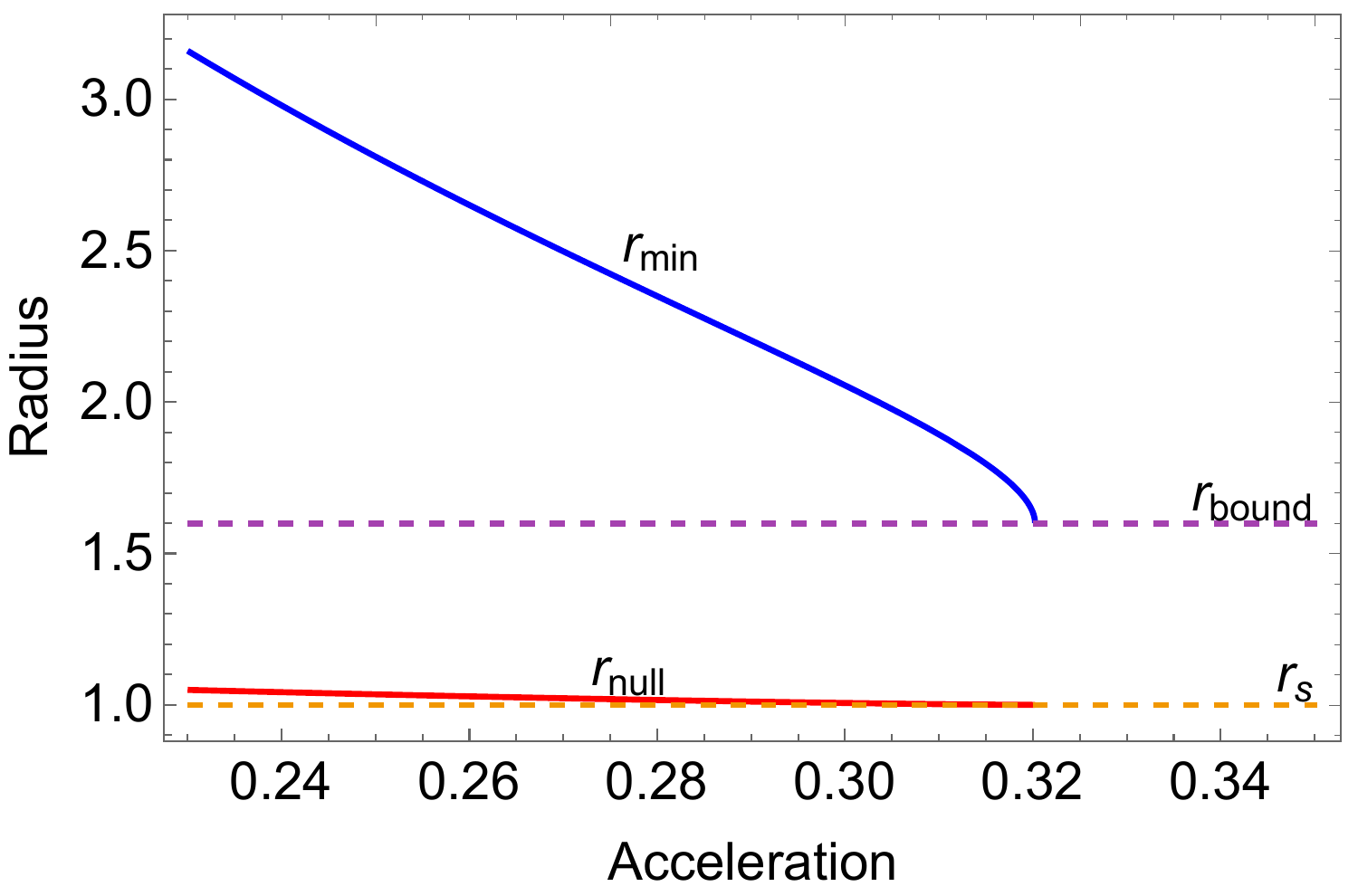}
\caption{h = 0.1}
\label{Bifurcation point for h=0.1}
\end{subfigure}
\begin{subfigure}{.5\textwidth}
\includegraphics[width=7.5cm, height=5cm]{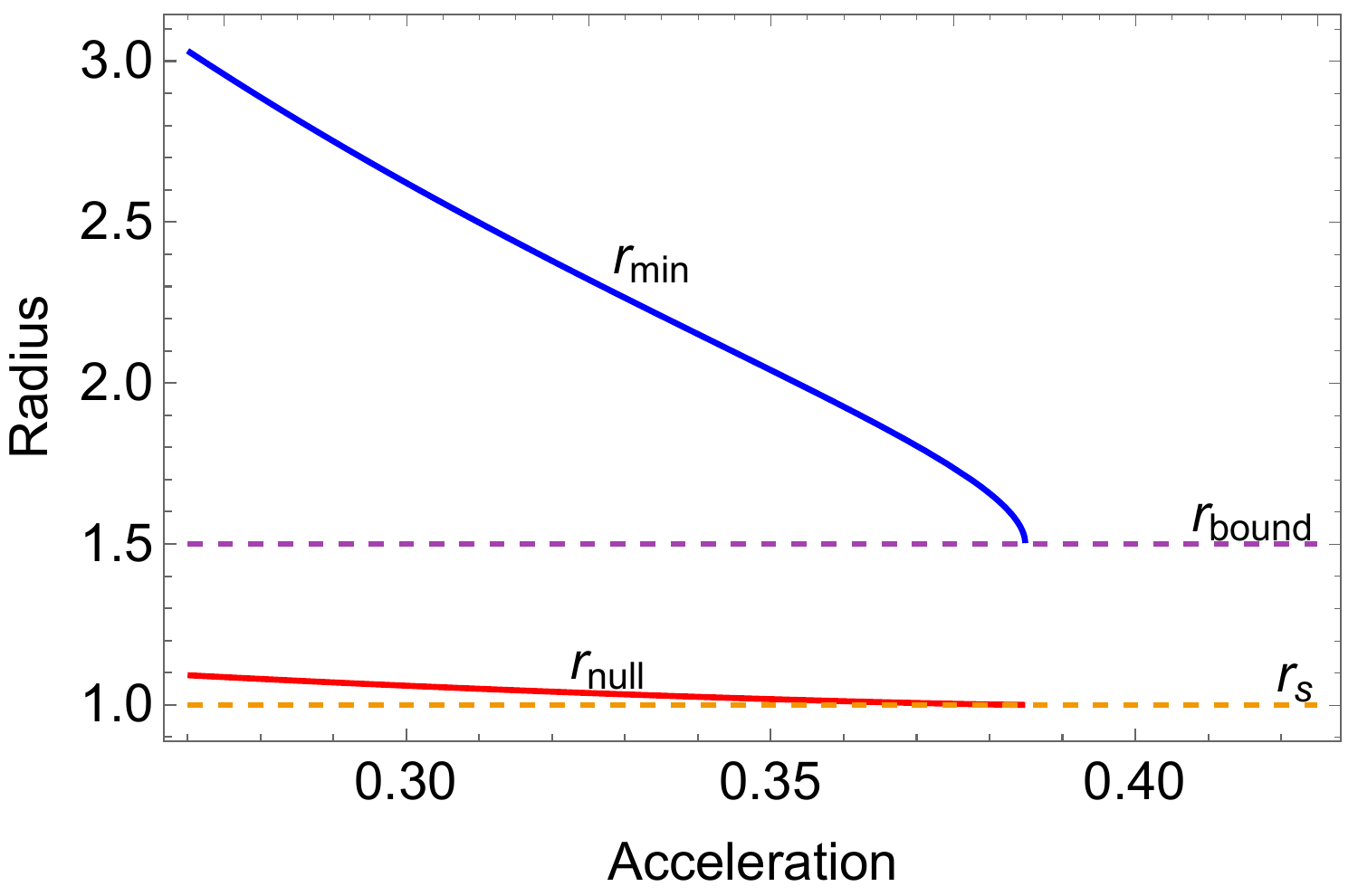}
\caption{h = 0}
\label{Bifurcation point for h=0}
\end{subfigure}
\begin{subfigure}{.5\textwidth}
\includegraphics[width=7.5cm, height=5cm]{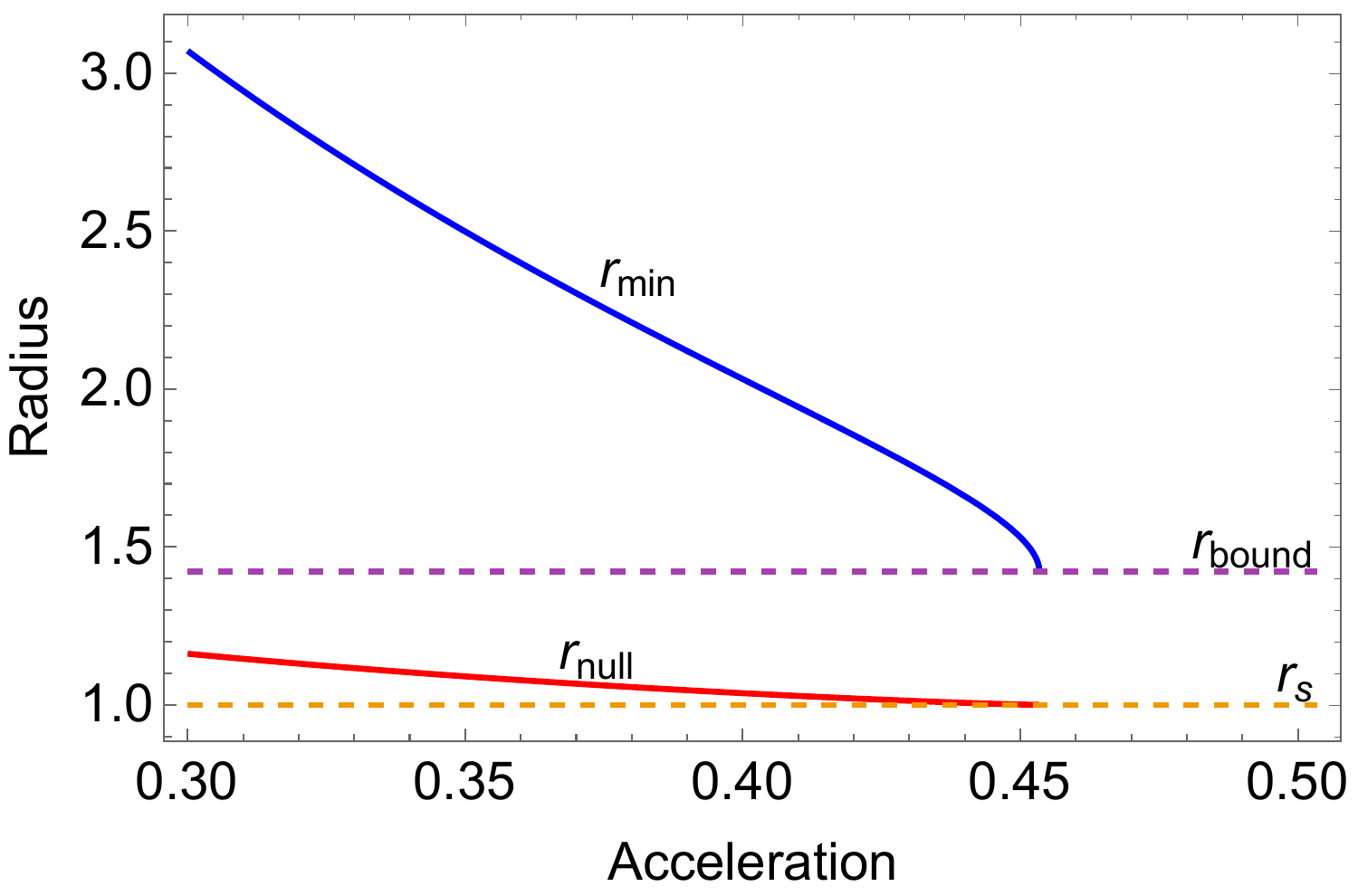}
\caption{h = -0.1}
\label{Bifurcation point for h=-0.1}
\end{subfigure}
\begin{subfigure}{.5\textwidth}
\includegraphics[width=7.5cm, height=5cm]{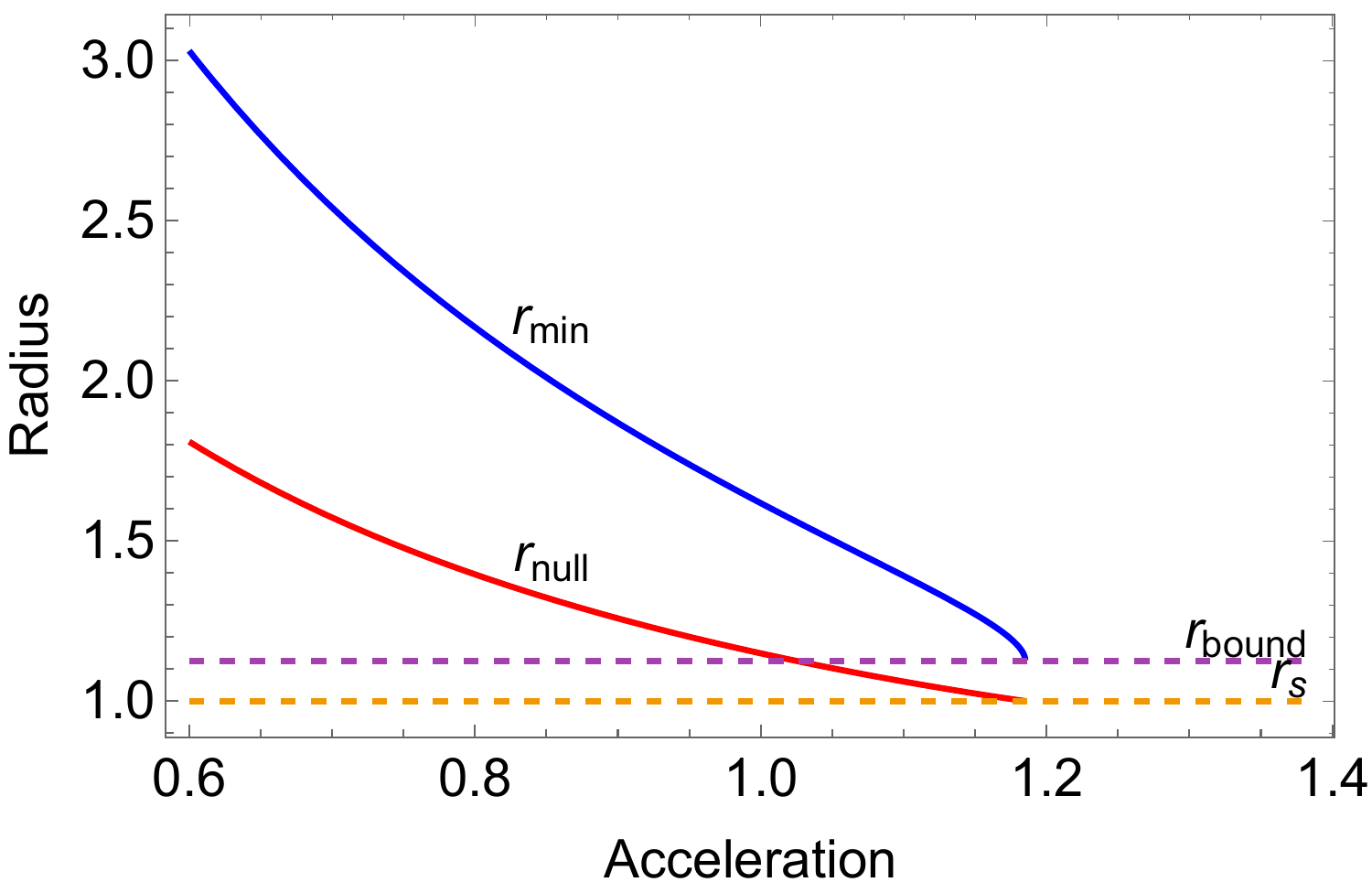}
\caption{h = -1}
\label{Bifurcation point for h=-1}
\end{subfigure}
\caption{Variation of $r_{null}$ and $r_{min}$ with magnitude of acceleration $|a|$ for four different values of $h$. The bounds on the value of acceleration $|a|_b$ are $0.3202$, $0.3849$, $0.4534$ and $1.1852$ for $h=0.1,0,-0.1$ and $-1$ respectively.}
\label{Bifurcation point plots}
\end{figure}
\begin{figure}[h]
\begin{subfigure}{.5\textwidth}
\includegraphics[width=7.25cm,height=5.25cm]{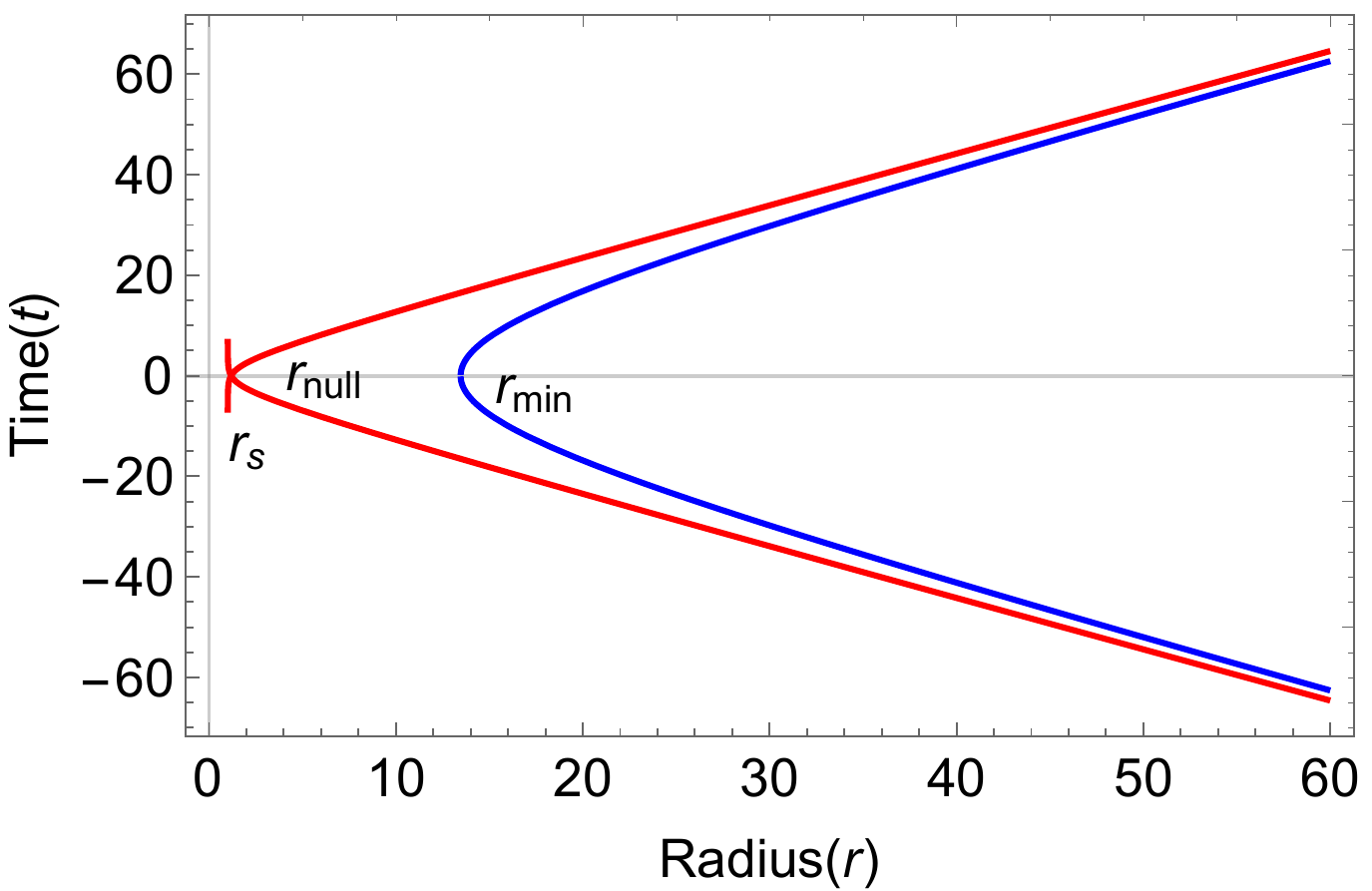}
\caption{$h = 0.1$}
\label{h=0.1}
\end{subfigure}
\begin{subfigure}{.5\textwidth}
\includegraphics[width=7.25cm,height=5.25cm]{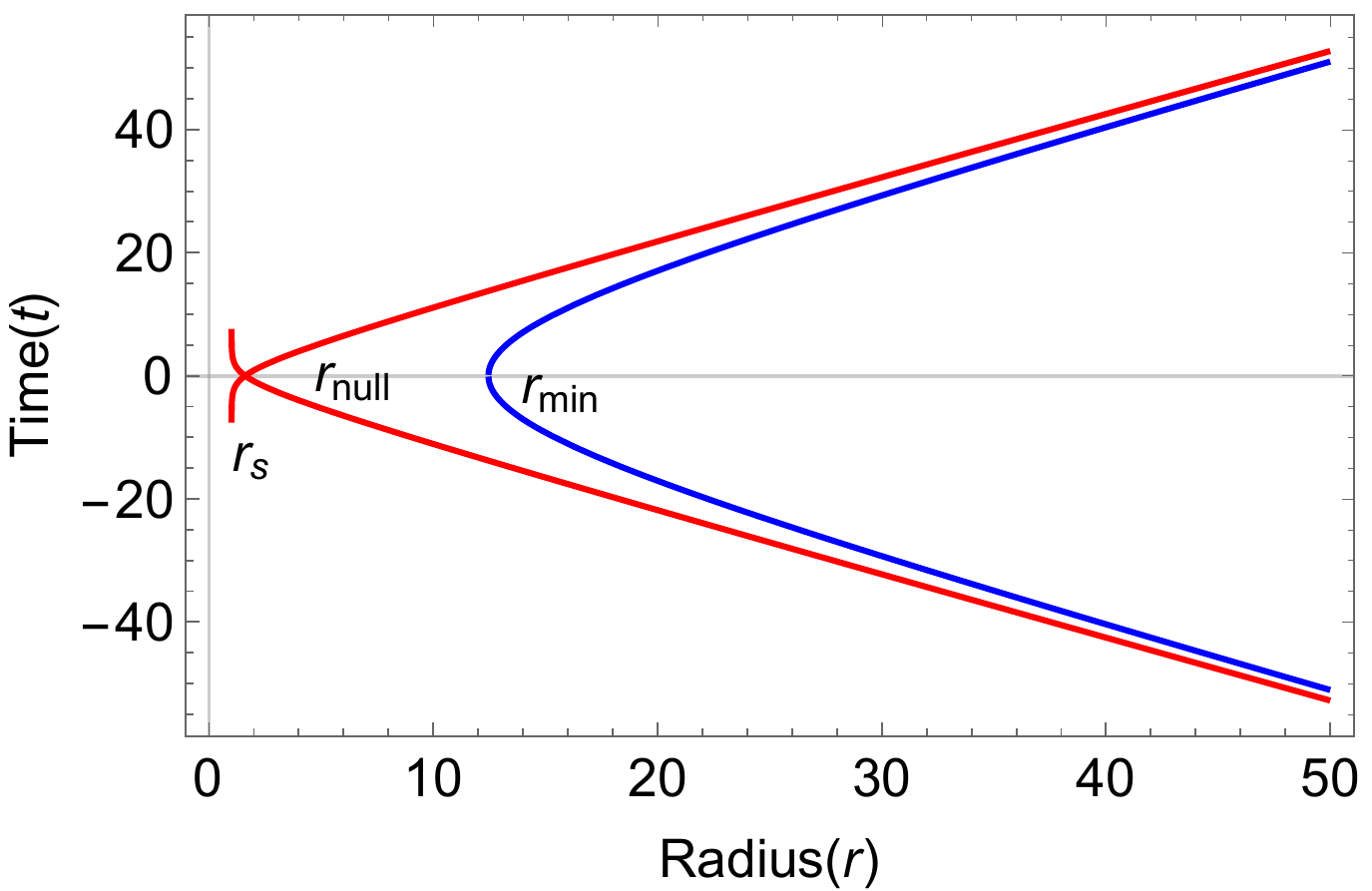}
\caption{$h = 0$}
\label{h=0}
\end{subfigure}
\begin{subfigure}{.5\textwidth}
\includegraphics[width=7.25cm,height=5.25cm]{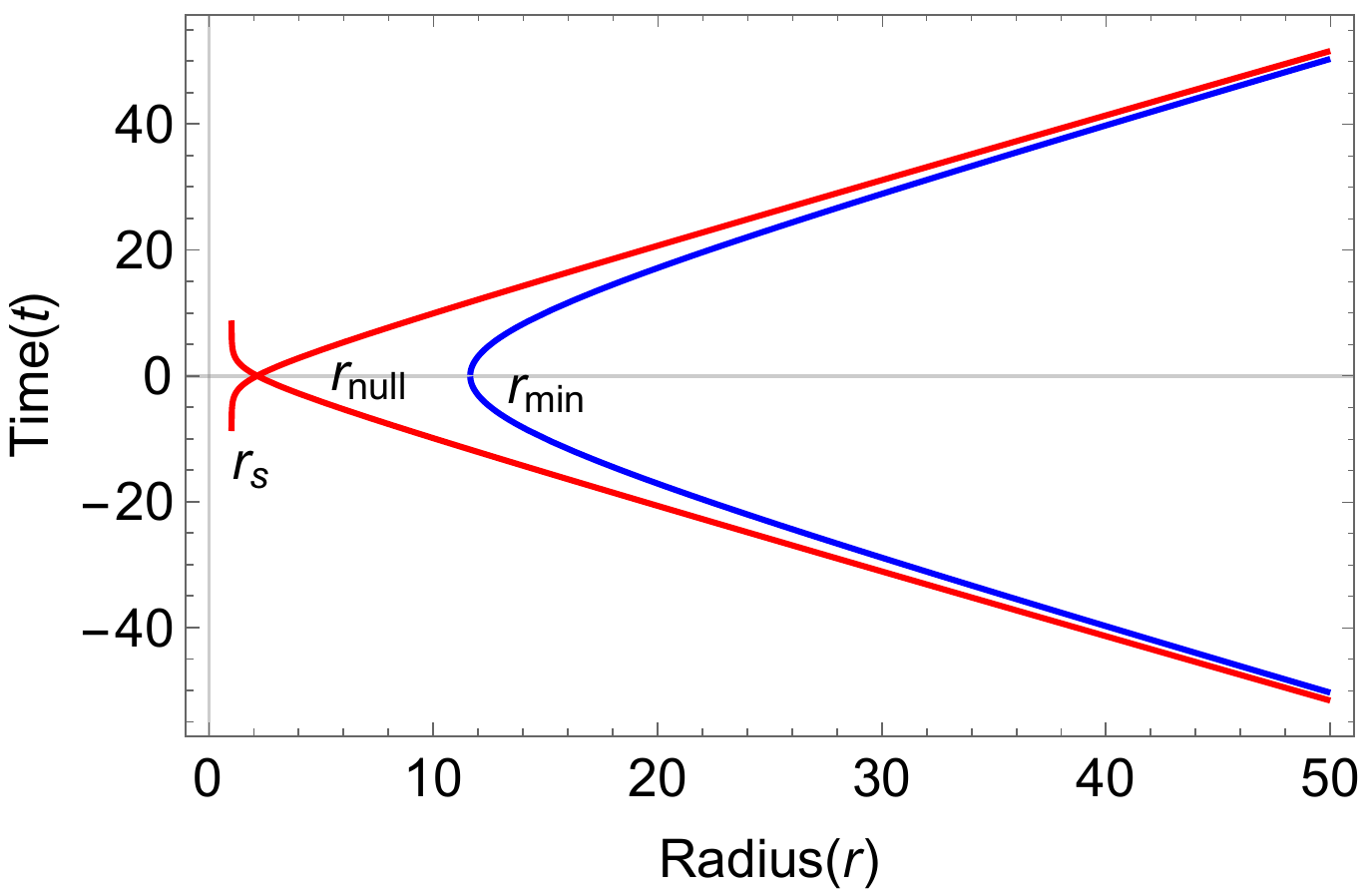}
\caption{$h = -0.1$}
\label{h=-0.1}
\end{subfigure}
\begin{subfigure}{.5\textwidth}
\includegraphics[width=7.25cm,height=5.25cm]{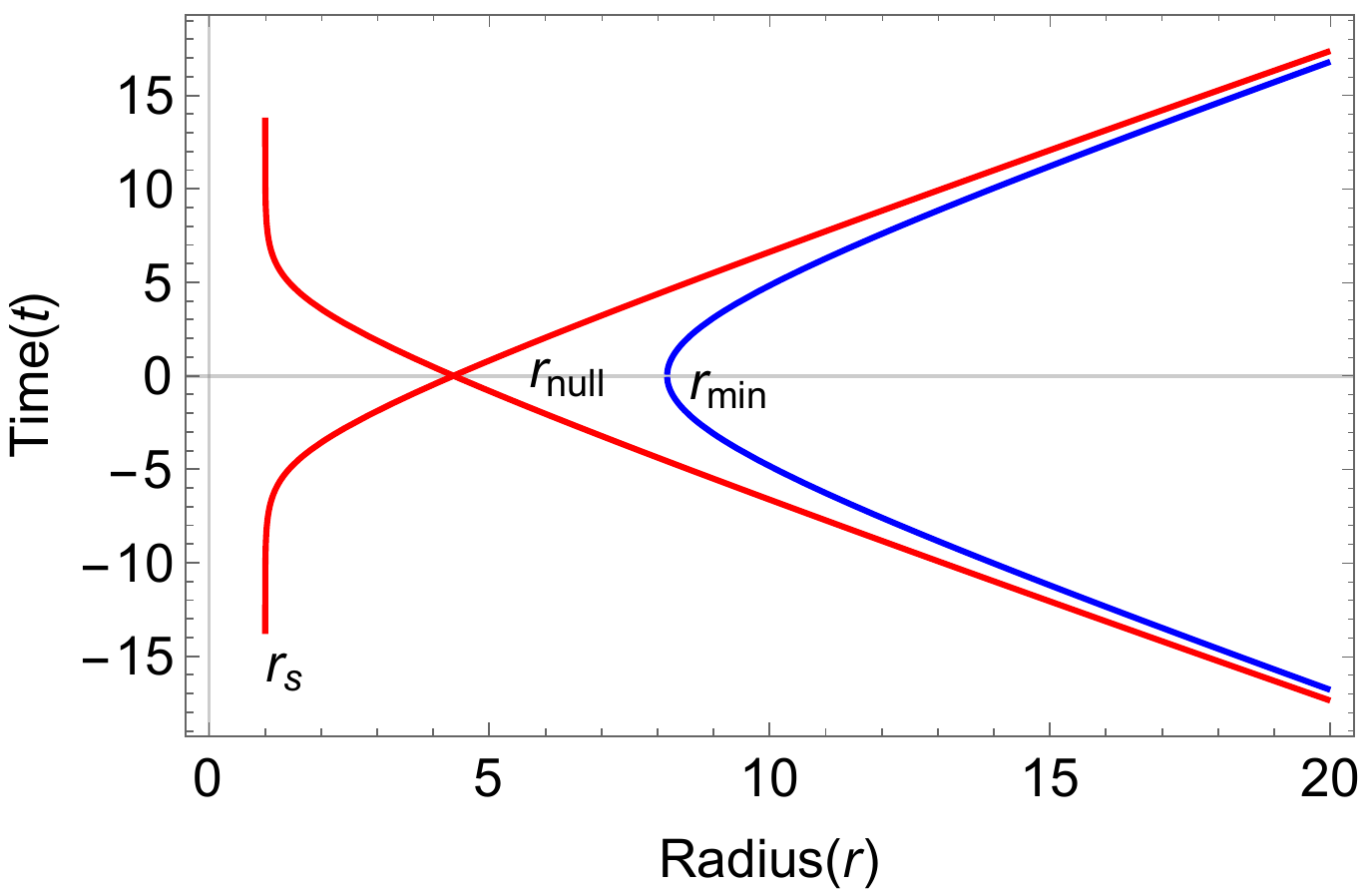}
\caption{$h = -1$}
\label{h=-1}
\end{subfigure}
\caption{Rindler horizons and LUA trajectories for four different values of initial data $h$ with acceleration $|a|={|a|}_b /5$ and $r_s =1$. }
\label{Rindler Quadrant}
\end{figure}

In Figure \ref{Bifurcation point plots}, we have plotted the distance of closest approach $r_{min}$ and the bifurcation point $r_{null}$  for some particular values of asymptotic initial data $h$.
(i) For $h \leq 0$ and for $h>0$ with $|a|>{|a|}_{min}$, increasing acceleration $|a|$ decreases both the distance of closest approach $r_{min}$ and the bifurcation point $r_{null}$. They approach the bound value $r_{b}$ and Schwarzschild radius $r_s$ respectively as $|a|$ approaches the bound value ${|a|}_b$.  Further, the radial difference, $r_{min} - r_{null}$ between the turning point $r_{min}$ and the bifurcation point $r_{null}$ decreases with increasing acceleration. The plots of Rindler quadrants and their respective LUA trajectories are shown in Figure \ref{Rindler Quadrant} for various values of asymptotic initial data $h$ and for a fixed value of acceleration $|a|={|a|}_b /5$ in each case. The corresponding values of $r_{min}$, $r_{null}$ and $r_{min}-r_{null}$ are given in the Table 1.  From the figures and the tabulated values, it is evident that shifting the trajectory closer to the black hole horizon by increasing the value of initial data $h$ accounts to increase the value of intercept $\cal C$, thereby decreasing the radial distance of the bifurcation point $r_{null}$ from black hole horizon at $r_s$. (ii) For $h>0$ with $|a|< {|a|}_{min}$, increasing $|a|$ decreases the value of $r_{min}$ while the $r_{null}$ increases upto its maximum value at $|a| = |a|_{min}$. The radial difference $r_{min}-r_{null}$ is again finite and non-zero in this case.

\begin{table}[h!]
\centering
\begin{tabular}{|c|c|c|c|c|}
\hline
$h$ & $|a|$ & $r_{min}$ & $r_{null}$ & $r_{min}-r_{null}$\\
\hline
$0.1$ & $0.06404$ & $13.4611$ & $1.18106$ & $12.28004$\\
\hline
$0$ & $0.07698$ & $12.4581$ & $1.61042$ & $10.84768$ \\
\hline
$-0.1$ & $0.09068$ & $11.646$ & $2.14481$ & $9.50119$ \\
\hline
$-1$ & $0.23704$ & $8.17092$ & $4.35977$ & $3.8115$ 
\\
\hline
\end{tabular}
\end{table}

These observations provide an alternate perspective to look at the acceleration bounds in Eq.(\ref{3.c- bound}) in the $(|a|, h)$ parametrisation. A comparison with the flat spacetime Rindler quadrant illustrates the case. In the flat spacetime, the acceleration of the Rindler trajectory can be increased all the way upto infinity, but still the trajectory is constrained to lie in the same Rindler quadrant. This is due to the reason that the corresponding intercept ${\cal C}$, the corresponding Rindler horizons including the bifurcation point are all independent of $|a|$. In fact, for the limiting case $|a| \rightarrow \infty$, the Rindler trajectory coincides with the past and future null horizon trajectories, that is, the turning point $r_{min}$ is then same as the bifurcation point. However, in the Schwarzschild case, the turning point $r_{min}$ can never be the same as the bifurcation point $r_{null}$ for finite asymptotic initial data $h$. This can be explained as follows: The intercept ${\cal C}$, the corresponding Rindler horizons and the bifurcation point are all functions of $|a|$ as well. Hence, increasing $|a|$ does decrease the turning point $r_{min}$, like in the flat Rindler case, however now the bifurcation point $r_{null}$ also decreases. The closest, a radial trajectory can be \textit{pushed}, say by increasing $|a|$ or varying $h$, towards the black hole is limited by how close the bifurcation point $r_{null}$ can get to the black hole. Here the lowest possible value $r_{null}$ can take is the Schwarzschild radius limited by the black hole horizon, which in turn limits the lowest possible value of $r_{min}$ to $r_{min} = r_b$ and hence a maximum value for the acceleration $|a|$. 
Pushing the LUA trajectory further towards the black hole, that is, by increasing $|a|$ and trying to lower $r_{min}$ than $r_b$, then figuratively, causes the corresponding $r_{null}$ to be inside the black hole horizon and hence no outgoing null geodesic exists which can reach the future null infinity ${\cal J^+}$ implying that there is no turning LUA trajectory in this case. Strictly, only for $|a| \rightarrow \infty$, can the LUA trajectory ever coincide completely with the past and future null horizon trajectories for the turning point $r_{min}$ to be same as the bifurcation point $r_{null}$ (which is true in the flat spacetime Rindler case only). Therefore, a difference $ r_{min} - r_{null} $ must always exist in the Schwarzschild case which implies the acceleration bound $|a|_b$ must exist.

\begin{figure}[h!]
\begin{subfigure}{.5\textwidth}
\includegraphics[width=7cm,height=5cm]{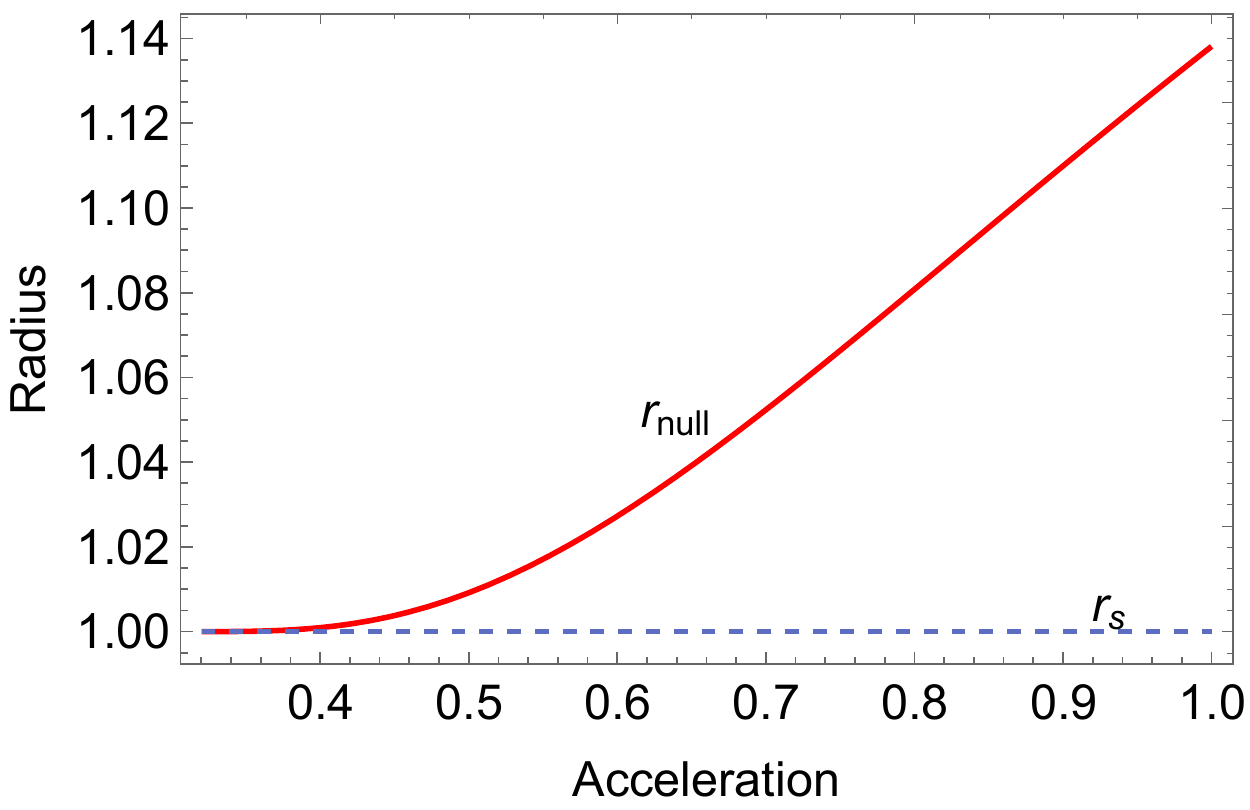}
\caption{$r_{min}=1.59772$}
\label{Bifurcation point for rmin=1.59772}
\end{subfigure}
\begin{subfigure}{.5\textwidth}
\includegraphics[width=7cm,height=5cm]{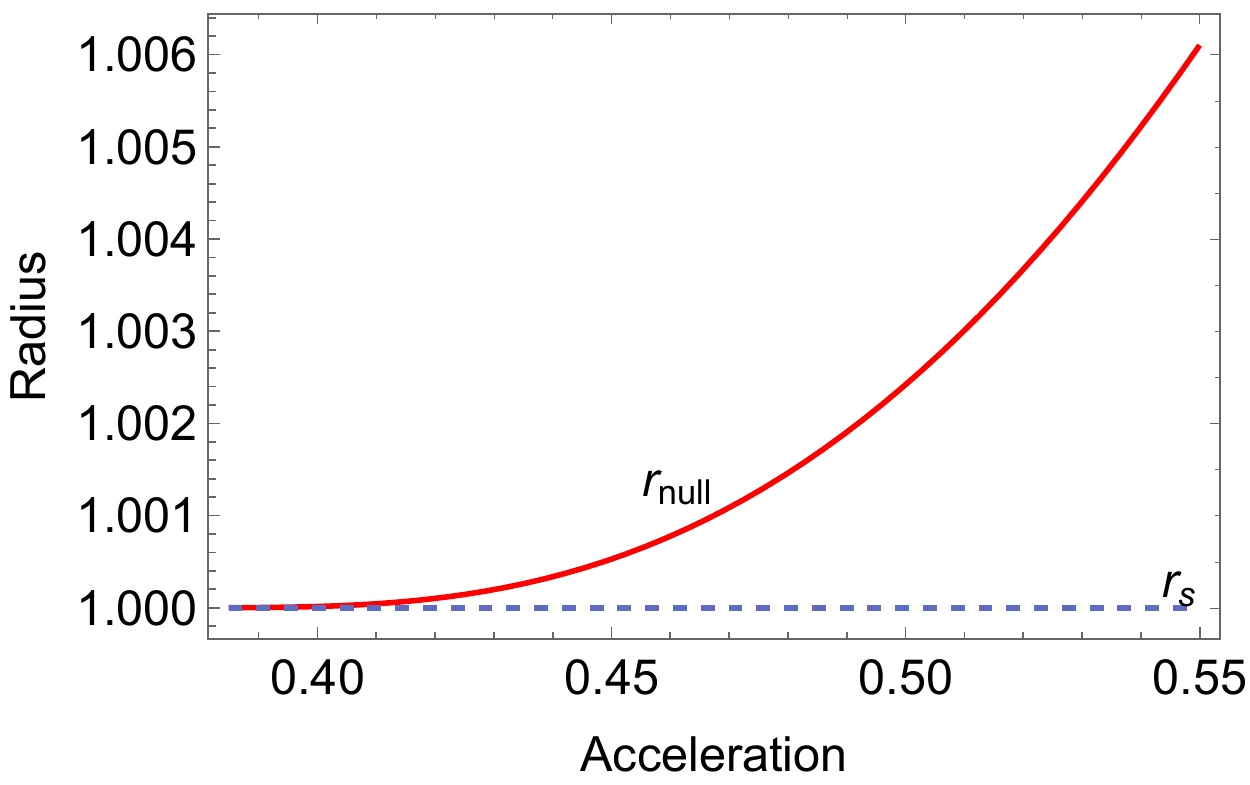}
\caption{$r_{min}=1.5$}
\label{Bifurcation point for rmin=1.5}
\end{subfigure}
\begin{subfigure}{.5\textwidth}
\includegraphics[width=7cm,height=5cm]{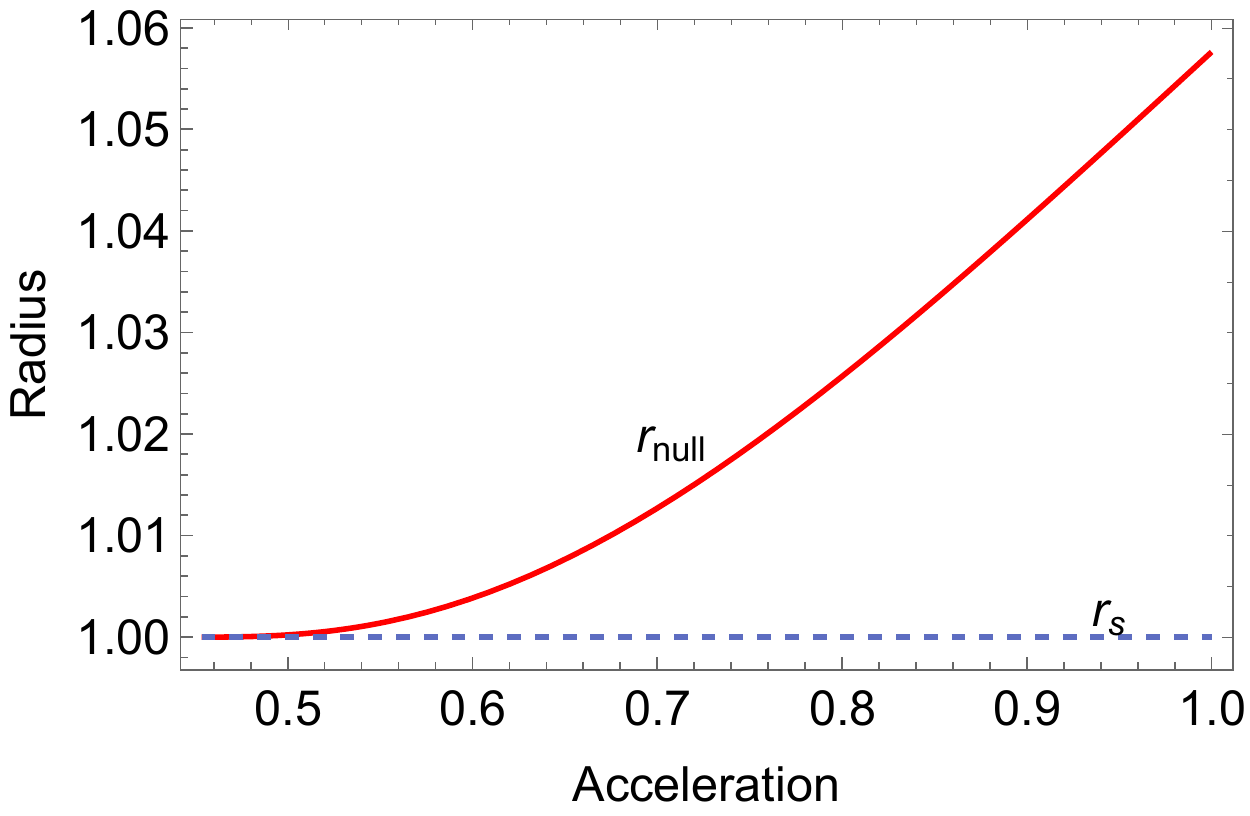}
\caption{$r_{min}=1.42248$}
\label{Bifurcation point for rmin=1.42248}
\end{subfigure}
\begin{subfigure}{.5\textwidth}
\includegraphics[width=7cm,height=5cm]{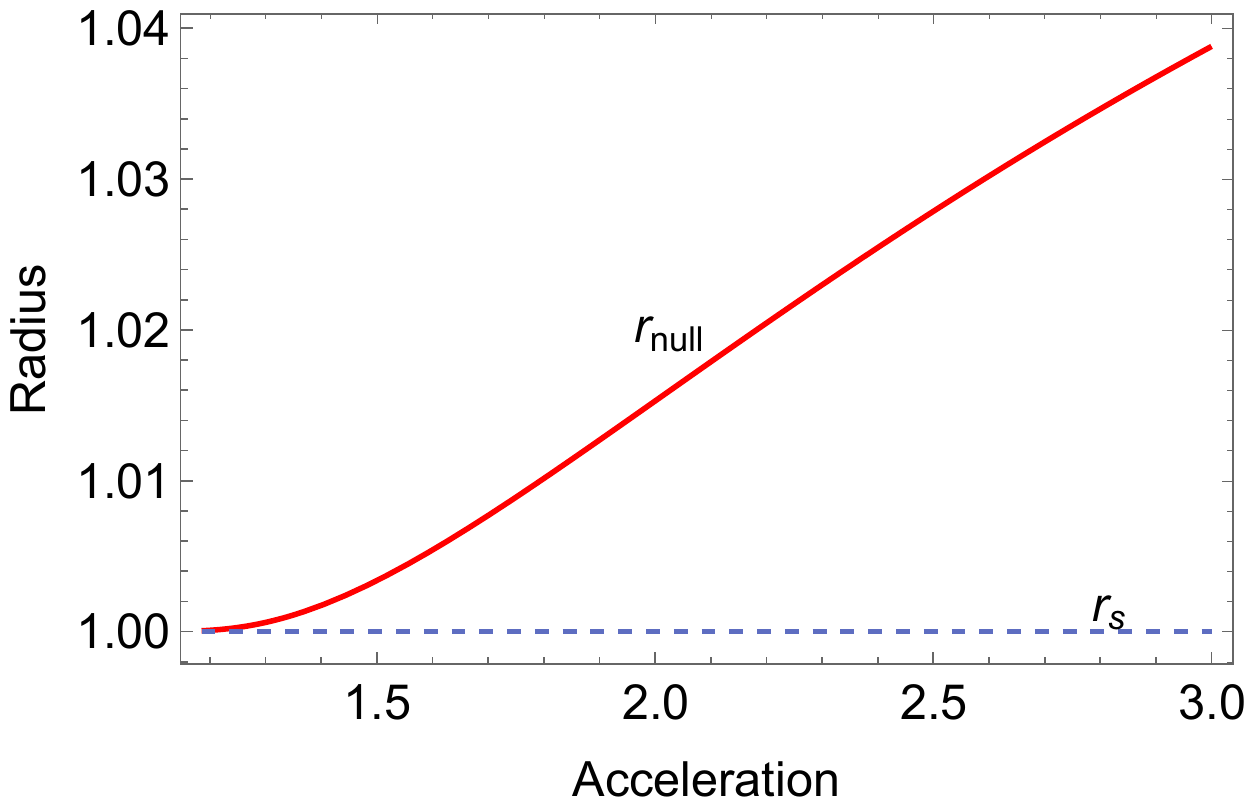}
\caption{$r_{min}=1.125$}
\label{Bifurcation point for rmin=1.125}
\end{subfigure}
\caption{Variation of bifurcation point $r_{null}$ with magnitude of acceleration $|a|$ for four different values of $r_{min}$.}
\label{Bifurcation point plots parametrization 2}
\end{figure}

In Figure \ref{Bifurcation point plots parametrization 2}, we have plotted the bifurcation point $r_{null}$ using the $(|a|,r_{min})$ parametrization, for particular values of turning point $r_{min}=1.59772,1.5,1.42248$ and $1.125$ with $r_s=1$. The lower bounds on the value of acceleration $|a|_b$ for these four values of $r_{min}$ are $0.3202$, $0.3849$, $0.4534$ and $1.1852$ respectively. The bifurcation point $r_{null}$ approaches the Schwarzschild radius $r_s$ as acceleration $|a|$ approaches the bound value $|a|_b$ as expected.

\subsection{Metric for the Rindler quadrant} \label{Metric}
Using the null geodesic reflection method of Bondi \cite{bondi}, it is possible to write down the form of the metric of the Rindler spacetime corresponding to a particular Rindler quadrant in the co-moving frame of the radial LUA observer as follows
\begin{eqnarray}
ds^2 &=&  \frac{f\left(U,V\right)}{f\left(U\right)f\left(V\right)} \left[\,|a| \, R(U) + h + \sqrt{\left[\;|a| \, R(V)+h\;\right]^2-f\left(V\right)}\,\right]\nonumber\\
& &  \left[\,|a| \, R(U) + h - \sqrt{\left[\;|a| \, R(U)+h\;\right]^2-f\left(U\right)}\,\right] \left(dU dV\right) \nonumber \\
& & -R^2 \left(d\theta^2 + \sin^2 \theta \, d\phi^2 \right) 
\label{lua metric}
\end{eqnarray} 
where $U=\tilde{t}-\tilde{r}$, $V=\tilde{t}+\tilde{r}$ are the null co-ordinates and $\tilde{t}$ and $\tilde{r}$ are the co-moving co-ordinates such that the LUA observer is at rest at $\tilde{r}=0$. Here $f(\tau) =  (1 - r_s/R(\tau))$ is the Schwarzschild metric function (or any general function $f(q)$ as defined in the metric in Eq.(\ref{general metric})) and $R(\tau)$ is the radial trajectory solution of Eq.(\ref{radial gen velocity}) as a function of the proper time $\tau$ along the trajectory. The explicit form of $R(\tau)$ is technically complex to write down but, in principle, one has to invert the expression of $\tau(R)$ in terms of elliptic integrals to arrive at one such analytical expression. The explicit co-ordinate transformation from the Schwarzschild co-ordinates $(t,r)$ to the co-moving co-ordinates $(\tilde{t},\tilde{r})$ can be found out from the following differential 
\begin{eqnarray}
dt &=& \frac{1}{2} \Bigg[ \left(\frac{u^1(V)}{f(V)}+u^0(V) \right)dV -\left(\frac{u^1(V)}{f(U)}-u^0(U) \right)dU\Bigg]\\
\frac{dr}{f(r)} &=& \frac{1}{2} \Bigg[ \left(\frac{u^1(V)}{f(V)}+u^0(V) \right)dV+\left(\frac{u^1(U)}{f(U)}-u^0(U) \right)dU\Bigg]
\end{eqnarray}
where the functions $u^0(\tau)$ and $u^1(\tau)$ are defined in Eqs.(\ref{temporal gen velocity}) and (\ref{radial gen velocity}) for the components of the four velocity of the LUA trajectory. At a large distance from the black hole $r\to\infty$, the spacetime is asymptotically flat that is $f(r)\to 1$ and the LUA trajectory takes the usual hyperbolic form of Rindler trajectory with $r = \left(\cosh(|a|\tau)-h\right)/|a|$ and $t = \sinh(|a|\tau)/|a|$. In this limit, the metric in Eq.(\ref{lua metric}) reduces to the usual form of the Rindler metric, 
\begin{eqnarray}
ds^2 &=&  \exp\left( 2 |a| \tilde{r} \right) \; \left(d{\tilde{t}}^2-d{\tilde{r}}^2\right)
\end{eqnarray}

The metric in Eq.(\ref{lua metric}) is in general dependent on the time co-ordinate $\tilde{t}$ as expected since the LUA observer is in motion with respect to the black hole and encounters different background curvature starting from zero value at $R \rightarrow \infty$ to its maximum encountered value at $R = r_{min}$. The future and past horizons correspond to $U \to\infty$ and $V\to-\infty$ curves respectively where the time-time component of the metric function vanishes and can be shown to lead to Eq.(\ref{future past Horizon}) for the null geodesics corresponding to the future and past Rindler horizons. Whereas, $U \to-\infty$ and $V\to +\infty$ correspond to past and future null infinity respectively.

\section{Discussion} \label{discussion}

The future and past intercept ${\cal C}$ of the radial LUA trajectory in a Schwarzschild spacetime with the future null infinity ${\cal J^+}$ and past null infinity ${\cal J^-}$ depends on both the magnitude of acceleration $|a|$ and the asymptotic initial data $h$, unlike in the flat Rindler spacetime case where it is only a function of translational shift $h$. The background curvature of the black hole not only affects the monotonicity of ${\cal C}$ due to $h$ but also initiates bounds on the values of the acceleration $|a|$ for the future Rindler horizon to exist. For a chosen Rindler quadrant, having a particular value of ${\cal C}$, there are infinitely many different combinations of $\{|a|, h \}_{{\cal C}}$ with each set having a different value of both $|a|$ and $h$ leading to the same intercept at the boundary. Furthermore, the turning point radius $r_{min}$ is different for each such set and the corresponding trajectories belonging to these sets do not overlap. Thus the set of all $\{|a|, h \}_{{\cal C}}$ trajectories with the same intercept ${\cal C}$ partially foliates the Rindler quadrant with the innermost trajectory, having the smallest $r_{min}$ close to the bound value $r_b$, being the inner boundary for the family of such curves. The region between the bifurcation point $r_{null}$ and $r_{b}$ is a no-go region for the turning LUA trajectories due to the acceleration bounds discussed in section \ref{acceleration bound}. Thus one could have a family of trajectories with constant acceleration ranging from zero to the maximum bound value $|a|_b$, which have the same Rindler horizon and hence a common Rindler quadrant, but each must have a different value of asymptotic initial data $h$. One caveat in the above discussion is that the analysis was restricted to the $\theta$ and $\phi$ constant hyperplane. The Rindler horizons in Eq.(\ref{future past Horizon}) were curves on the null surface which form the horizons. These would require a careful analysis of null geodesics which can travel in the transverse directions as well. Nevertheless, the analysis presented in the paper is an outset of a much required broader analysis in the $3+1$ case while it is complete for the $1+1$ dimension black hole.   

Investigating quantum field effects perceived by the LUA observers in the Boulware and Hartle-Hawking states of the black hole would be the next interesting question to probe in the present context. The radially moving LUA observer in the Schwarzschild spacetime is analogous to a Rindler observer moving in an existing thermal bath, which in the present case is the Hawking thermal bath of the black hole corresponding to the Hartle-Hawking state. The two scales involved, the Hawking temperature inversely proportional to the mass of the black hole and the acceleration scale $|a|$ of the Unruh bath temperature, would both be relevant in such an investigation. In an earlier work \cite{kolekar2, kolekar3}, the quantum field aspects for a uniformly accelerated observer moving in an inertial thermal bath were investigated. Here too, there are two temperature scales, the inertial thermal bath with temperature $T_{b}$ and $T_{u} = |a|/(2 \pi)$, the Rindler horizon temperature. It was shown that the reduced density matrix for the Rindler observer in a flat spacetime moving in an inertial thermal bath (instead of the usual inertial vacuum) with acceleration $|a| = 2 \pi T_{u}$, is symmetric in $T_{u}$ and $T_{b}$. It was argued that the Rindler observer is unable to distinguish between thermal and quantum fluctuations. It would be interesting to check whether a similar indistinguishability holds even in the Schwarzschild case.

\section*{Acknowledgments}

SK and KP thank the Department of Science and Technology, India, for financial support.

\end{document}